\documentclass[submitting]{nst}

\usepackage{dcolumn}
\usepackage[T2A,T1]{fontenc}
\usepackage[russian,english]{babel}
\usepackage{multirow}
\usepackage{listings}
\usepackage{xcolor}
\usepackage[caption=false]{subfig} 
\begin{document}

\title{The physical design of the superfluid $^{4}\mathrm{He}$ UCN source at CSNS}

\author{Xue-Fen Han}
\affiliation{Institute of High Energy Physics, Chinese Academy of Sciences (CAS), Beijing 100049, China}
\affiliation{Spallation Neutron Source Science Center (SNSSC), Dongguan 523803, China}
\affiliation{University of Chinese Academy of Sciences, Beijing 100049, China}
\affiliation{Department of Physics, Technical University of Munich, 85748 Garching, Germany}
\affiliation{Heinz Maier-Leibnitz Zentrum, Technical University of Munich, Lichtenbergstra\ss{}e 1, 85748 Garching, Germany}

\author{Bin Zhou}
\affiliation{Institute of High Energy Physics, Chinese Academy of Sciences (CAS), Beijing 100049, China}
\affiliation{Spallation Neutron Source Science Center (SNSSC), Dongguan 523803, China}

\author{Song-Lin Wang}
\affiliation{Institute of High Energy Physics, Chinese Academy of Sciences (CAS), Beijing 100049, China}
\affiliation{Spallation Neutron Source Science Center (SNSSC), Dongguan 523803, China}

\author{Xiao-Xiao Cai}
\affiliation{Institute of High Energy Physics, Chinese Academy of Sciences (CAS), Beijing 100049, China}
\affiliation{Spallation Neutron Source Science Center (SNSSC), Dongguan 523803, China}

\author{Dong-Dong Zhou}
\affiliation{Institute of High Energy Physics, Chinese Academy of Sciences (CAS), Beijing 100049, China}
\affiliation{Spallation Neutron Source Science Center (SNSSC), Dongguan 523803, China}

\author{Yi-Jie Cai}
\affiliation{Institute of High Energy Physics, Chinese Academy of Sciences (CAS), Beijing 100049, China}
\affiliation{Spallation Neutron Source Science Center (SNSSC), Dongguan 523803, China}

\author{Fei Shen}
\email[Corresponding author, shenf@ihep.ac.cn]{1 Zhongziyuan Road, Dongguan 523803, 15611938223}
\affiliation{Institute of High Energy Physics, Chinese Academy of Sciences (CAS), Beijing 100049, China}
\affiliation{Spallation Neutron Source Science Center (SNSSC), Dongguan 523803, China}

\author{Tian-Jiao Liang}
\email[Corresponding author, liangtj@ihep.ac.cn]{1 Zhongziyuan Road, Dongguan 523803, 15611938223}
\affiliation{Institute of High Energy Physics, Chinese Academy of Sciences (CAS), Beijing 100049, China}
\affiliation{Spallation Neutron Source Science Center (SNSSC), Dongguan 523803, China}
\affiliation{University of Chinese Academy of Sciences, Beijing 100049, China}
\affiliation{Guangdong-Hong Kong-Macao Joint Laboratory for Neutron Scattering Science and Technology, Dongguan 523803, Guangdong, China}

\author{Robert Golub}
\affiliation{Department of Physics, North Carolina State University, Raleigh, NC 27695, USA}

\begin{abstract}

Ultracold neutrons (UCN) play a crucial role in fundamental physics research, enabling investigations into new sources of CP violation, neutron lifetime measurements, and the exploration of non-Newtonian forces. However, these experiments are often limited by the intensity of UCN sources. The China Spallation Neutron Source (CSNS) is developing a UCN source utilizing superfluid $^{4}\mathrm{He}$ at Beamline 12. This paper focuses on enhancing UCN density and total UCN number by analyzing various parameters that impact the UCN system, including the influence of different wavelengths of cold neutrons, incident beam size, converter length, wall and reflector materials, extraction method, extraction guide radius, and guide material composition. A preliminary design of the UCN source at CSNS is presented based on this analysis. For a Be-coated converter, the predicted saturated UCN density is 315~UCN/cm$^3$, with a UCN production rate of 12,055~UCN/s and an extraction efficiency of 84\%, yielding approximately $1.3\times10^6$ extracted UCNs per cycle in accumulation mode. The UCN source design presented in this study forms part of the overall physical design of the Neutron Physics and Application Spectrometer at CSNS, which successfully passed its physical design review on March 25, 2025.

\end{abstract}


\maketitle
\nolinenumbers

\section{Introduction}\label{sec1}

Slow neutrons undergo total reflection when they collide with the surface of most materials at glancing angles. When their velocity is sufficiently low, the material atoms appear as a constant average potential rather than individual nuclei with strong potentials \cite{bib:1}. These ultracold neutrons can be totally reflected at any incident angle and can be stored for a long time, allowing measurements of neutron properties. This feature has fostered their application in a range of fundamental physics experiments, including neutron electric dipole moment (nEDM) measurements \cite{bib:2}, neutron lifetime determinations \cite{bib:3}, beta decay studies \cite{bib:4,bib:5}, and searches for deviations from Newton's gravity law at the micrometer length scale \cite{bib:6,bib:7}. However, the practical implementation of these experiments is often constrained by the available UCN intensity within the storage volume. Thus, developing strong UCN sources is essential.

In 1969, the first experimental observations of UCNs were reported independently by Shapiro et al. at Dubna \cite{bib:8} and by A. Steyerl in Munich \cite{bib:9}. At the VVR-M reactor of the Leningrad Institute of Nuclear Physics (now the Petersburg Nuclear Physics Institute, PNPI), a universal source employing a liquid-hydrogen moderator in the reactor core delivered intense beams of cold, polarized cold, and ultracold neutrons \cite{bib:10}. At the Institut Laue-Langevin (ILL), very-cold neutrons from a liquid-deuterium cold source were transported through a vertical guide system to the Steyerl neutron turbine, where multiple total reflections from moving curved blades Doppler-shifted the neutrons into the UCN energy range \cite{bib:11,bib:12}. In a thermal-equilibrium Maxwellian spectrum, UCNs constitute only about $10^{-12}$--$10^{-11}$ of the thermal-neutron flux \cite{bib:13}. Although gravitational and mechanical deceleration can shift neutrons into the UCN energy range, Liouville's theorem requires such conservative transformations to preserve neutron phase-space density, thereby imposing a fundamental limit on the UCN phase-space density attainable with sources based solely on moderation and spectral transformation \cite{bib:14}.

In 1975, R. Golub and Pendlebury proposed the concept of a ``superthermal UCN source'' to enhance UCN density by utilizing inelastic neutron scattering mechanisms more efficiently than existing UCN sources \cite{bib:15}. By 1977, they recommended the use of superfluid $^{4}\mathrm{He}$ as the UCN source material \cite{bib:16}. In 1983, R. Golub and K. Böning suggested employing a thin layer of solid deuterium as an alternative UCN source material \cite{bib:17}. Currently, superfluid $^{4}\mathrm{He}$ and solid deuterium are the most commonly utilized converter materials for superthermal UCN sources. Operational UCN sources based on superfluid $^{4}\mathrm{He}$ include SUN1 \cite{bib:18,bib:19}, SUN2 \cite{bib:20} and SuperSUN \cite{bib:21} at ILL, and TRIUMF \cite{bib:22,bib:14}. PNPI \cite{bib:23} is constructing superfluid $^{4}\mathrm{He}$ UCN sources. Solid deuterium UCN sources at the Paul Scherrer Institute (PSI) \cite{bib:24,bib:25,bib:26}, Los Alamos National Laboratory (LANL) \cite{bib:27}, and TRIGA Mainz \cite{bib:28} are operational, while those at FRM II \cite{bib:29} and PULSTAR \cite{bib:30} are currently under construction. Additionally, superfluid $^{4}\mathrm{He}$ UCN source concepts are under development at the European Spallation Source (ESS) \cite{bib:31,bib:32} and at the WWR-K research reactor of the Institute of Nuclear Physics ( INP ) in Almaty, Kazakhstan (AlSUN) \cite{bib:33,bib:34,bib:35}.

In terms of their implementation, UCN sources can be classified into two main types: in-pile and in-beam sources. In-pile UCN sources, located close to a spallation target station or a reactor core, can use either superfluid $^{4}\mathrm{He}$ or solid deuterium as the converter material. 
In contrast, in-beam UCN sources, situated far away from the target station or the reactor core, typically use superfluid $^{4}\mathrm{He}$ as the converter material. This preference stems from the fact that solid deuterium UCN sources of the in-beam type must have a relatively small volume to overcome the challenge of extracting UCNs from the bulk of the solid deuterium converter. This limitation can result in a lower overall UCN production rate \cite{bib:31,bib:32}.
Thus, the UCN source based on Beamline 12 at CSNS is an in-beam source, specifically a superfluid $^{4}\mathrm{He}$ UCN source \cite{bib:36}.

In this study, a preliminary physical design for the CSNS UCN source is established through a comprehensive analysis of parameters governing UCN density, production, and extraction efficiency. Section II outlines the source infrastructure, detailing the neutron flux from various moderators and the layout of Beamline 12. Section III presents the methodology and simulation framework employed for the source design. MCNPX is used to simulate cold neutron transport in the superfluid $^{4}\mathrm{He}$ converter \cite{bib:37}, while iMCUCN is employed to simulate the subsequent UCN storage and extraction processes \cite{bib:36,bib:38}. Section IV systematically investigates the key factors influencing the performance of the superfluid $^{4}\mathrm{He}$ converter at CSNS. Finally, Section V specifies the technical parameters and predicted performance of the initial version of the CSNS UCN source.

\section{UCN source at CSNS}\label{sec2}

The \SI{62.5}{\micro\ampere} beam of 1.6 GeV protons produced by the CSNS accelerator complex impinges on the tungsten spallation target. To further enhance research capabilities, the CSNS-II upgrade aims to achieve a design beam power of 500 kW \cite{bib:39,bib:40,bib:41,bib:42,bib:43}. This significant power increase provides a robust foundation for high-intensity neutron applications, including the development of advanced spectrometers at Beamline 12.

\begin{figure}[htbp]
    \centering
    \includegraphics[width=0.45\textwidth]{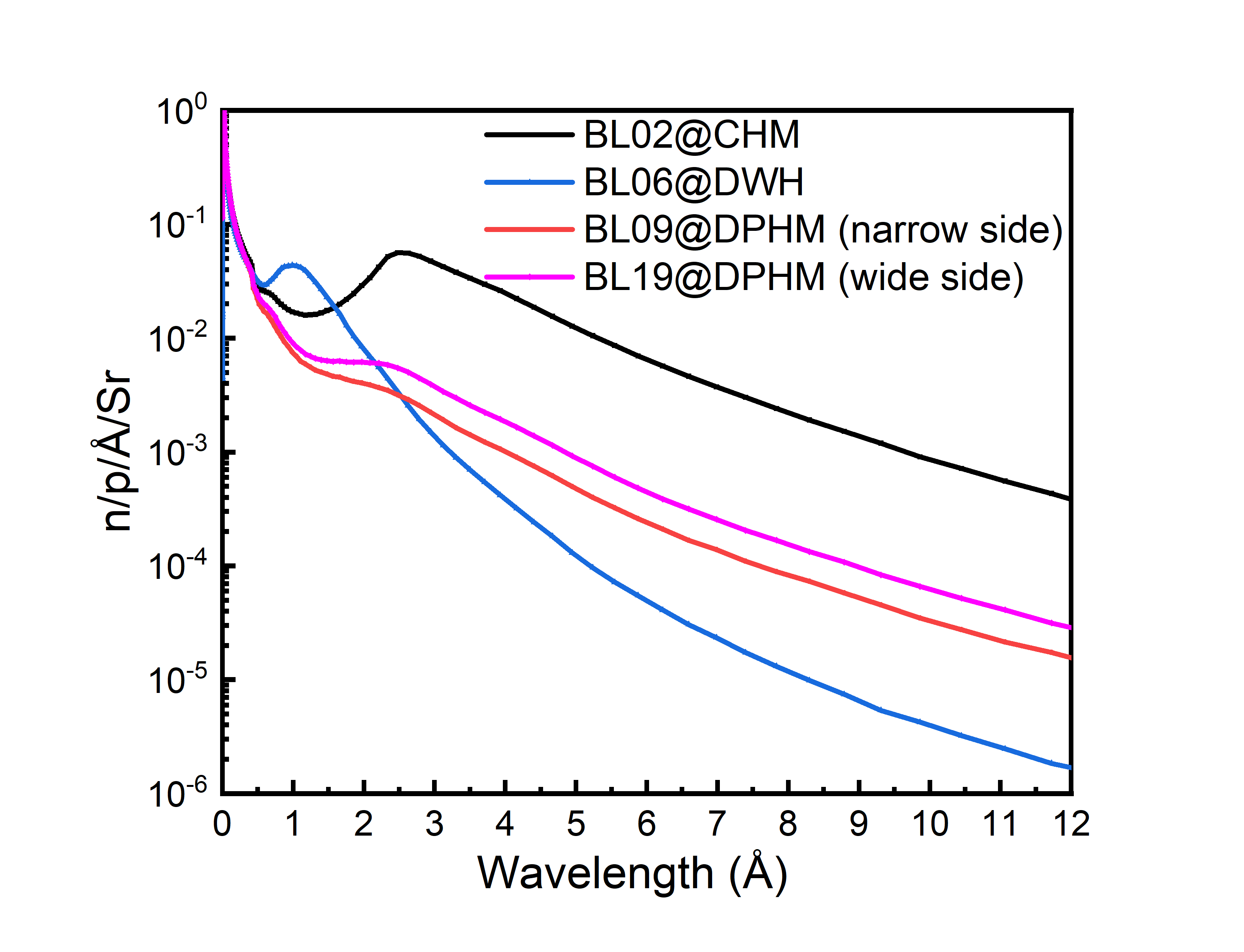}
    \caption{The neutron flux of different moderators at CSNS.}
    \label{Moderator}
\end{figure}

Superfluid  $^{4}\mathrm{He}$ UCN sources primarily require cold neutrons with wavelengths ranging from 4 to 10 Å. Figure \ref{Moderator} illustrates the neutron wavelength spectra emitted by three types of moderators at CSNS. Among these, the coupled hydrogen moderator (CHM) exhibits the highest cold neutron intensity, approximately an order of magnitude greater than that of either of the other two moderators, making it the most suitable cold neutron source for UCN applications.

Beamline 12 will be divided into two sub-beamlines supplied by the CHM. One scattering room will feature prompt gamma analysis and neutron depth profiling devices, while the other will host China's first UCN source.  Following moderation in the coupled liquid hydrogen moderator, a fraction of the cold neutrons will be directed towards the superfluid $^{4}\mathrm{He}$ UCN source through a curved neutron guide, which effectively filters out fast neutrons and high-energy gamma radiation to optimize the signal-to-background ratio.

\section{Calculation and simulation method}\label{sec:calculation-simulation}

\begin{figure*}[htbp]
    \centering
    \subfloat[Cold-neutron transport and UCN production.]
    {
        \includegraphics[width=0.48\textwidth]{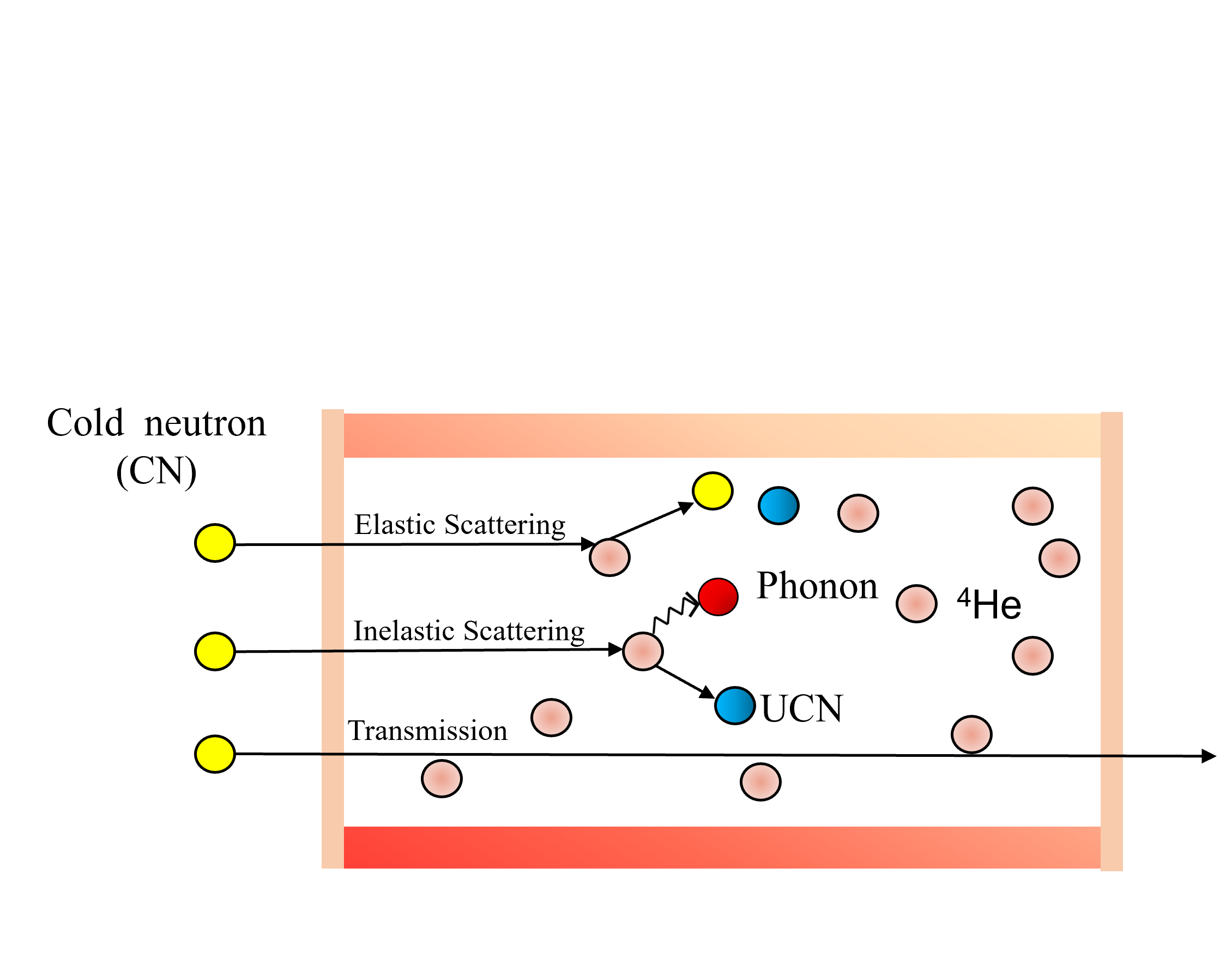}
        \label{fig:CNphysics}
    }
    \hfill
    \subfloat[UCN storage, loss, and extraction.]
    {
        \includegraphics[width=0.48\textwidth]{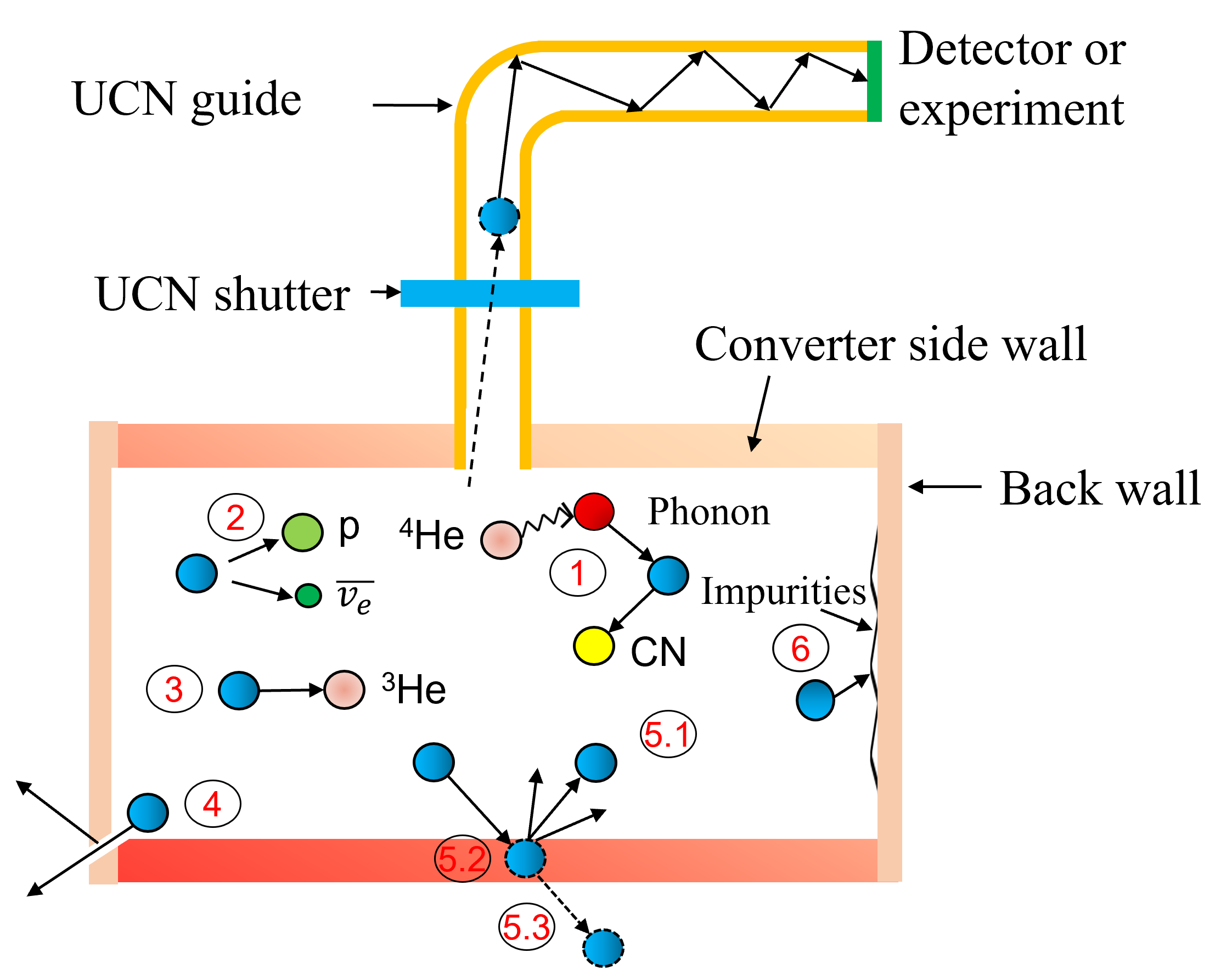}
        \label{fig:UCNPhysics}
    }
    \caption{Physical processes involved in the superfluid $^{4}\mathrm{He}$ UCN source at CSNS.}
    \label{fig:source-physics}
\end{figure*}

To evaluate and optimize the CSNS UCN source, the computational workflow combines MCNPX simulations of cold-neutron transport, numerical calculations of UCN production, and iMCUCN simulations of UCN storage and extraction. The physical processes represented in this workflow are summarized in Fig.~\ref{fig:source-physics}.

As illustrated in Fig.~\ref{fig:CNphysics}, most incident cold neutrons pass through the superfluid $^{4}\mathrm{He}$ converter, whereas a fraction interact with the converter medium. Only a small fraction lose sufficient energy through single-phonon emission by incident neutrons near 8.9~\text{\AA} or through multiphonon emission by incident neutrons in the 4--8~\text{\AA} range, and are thereby converted into UCNs.

As shown in Fig.~\ref{fig:UCNPhysics}, the produced UCNs can be stored in the converter but may be lost through several mechanisms, including upscattering to cold-neutron energies, absorption by $^{3}\mathrm{He}$, absorption or inelastic scattering at the converter walls, leakage through gaps, $\beta$ decay, and other secondary processes. The UCN population ultimately approaches a dynamic equilibrium between production and loss, after which the stored UCNs can be extracted through the UCN guide.

\subsection{Cold neutron transport}\label{subsec:cold-neutron-transport}

\begin{figure}[h]%
\centering
\includegraphics[width=0.45\textwidth]{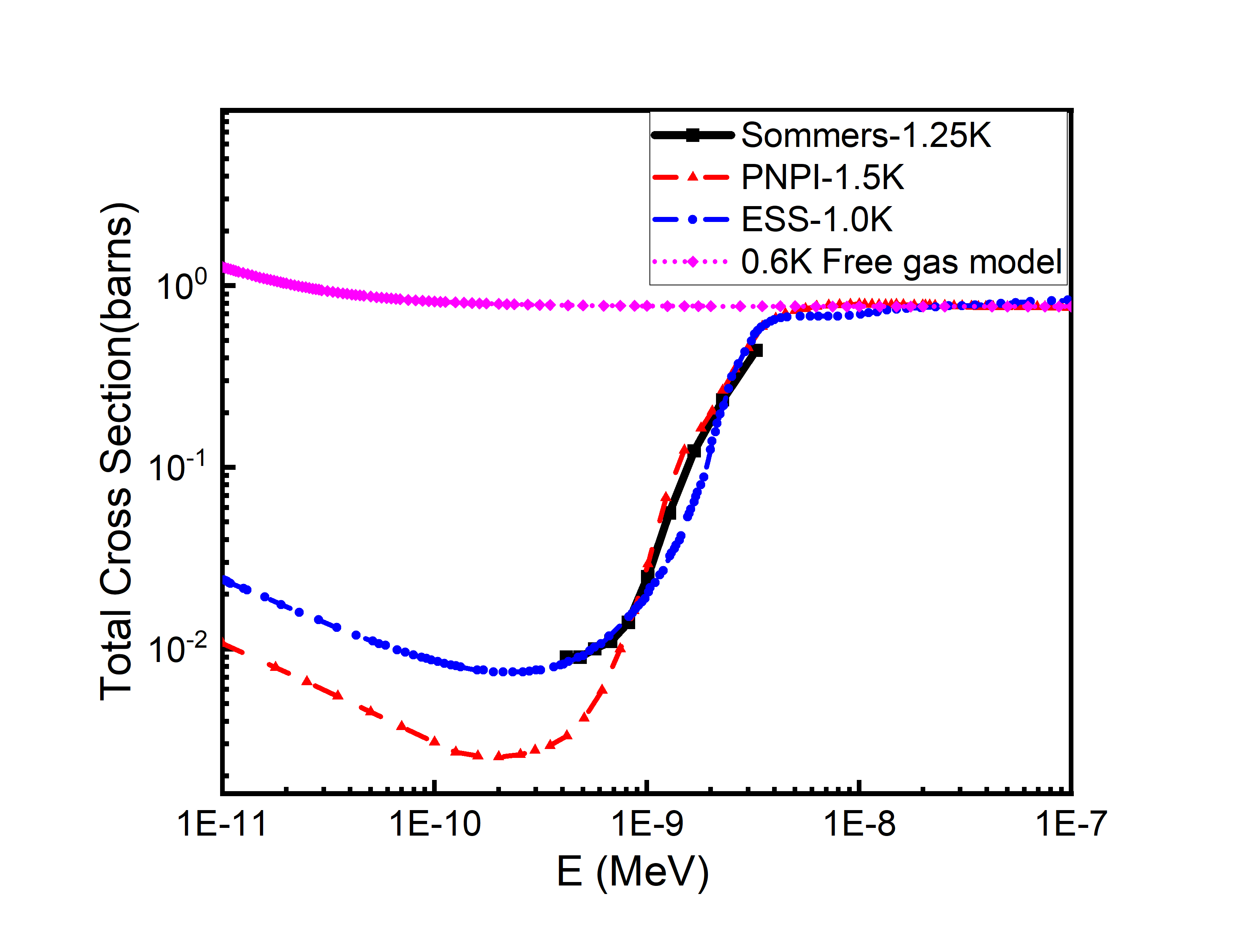}
\caption{Comparison of cold neutron cross sections for superfluid $^{4}\mathrm{He}$: experimental data measured by Sommers et al. at 1.25 K \cite{bib:44}, thermal scattering cross sections at 1.5 K (PNPI) \cite{bib:45} and from ESS \cite{bib:46}, and the 0.6 K cross section calculated via the free gas model using NJOY \cite{bib:47}.}\label{cold_He_cs}
\end{figure}

MCNPX 2.5.0 \cite{bib:37} is used to simulate cold neutron transport in superfluid $^{4}\mathrm{He}$. The cold neutron cross section of superfluid $^{4}\mathrm{He}$ plays a crucial role in the cold neutron transport stage. Fig. \ref{cold_He_cs} illustrates different types of cold neutron cross sections for superfluid $^{4}\mathrm{He}$, including the 1.25 K cross section measured by Sommers \cite{bib:44}, the 1.5 K thermal scattering cross section provided by Dr. M. S. Onegin of PNPI \cite{bib:45}, the thermal scattering cross section reported by ESS \cite{bib:46}, and the 0.6 K cross section generated using the free gas model with NJOY 2016.71 \cite{bib:47}. For pure superfluid $^{4}\mathrm{He}$, neutron absorption is negligible, and the thermal scattering models considered here contain no separate elastic scattering contribution. Therefore, the total neutron cross section is represented by the inelastic scattering cross section, which includes both single-phonon and multiphonon processes \cite{bib:46}.

The primary neutrons responsible for UCN production are cold neutrons with wavelengths ranging from 4 to 10 Å, corresponding to neutron energies of approximately 0.82–5.11 meV. Within this range, the 1.5 K thermal scattering cross section provided by Dr. Onegin Mikhail is closest to the cross-section data measured by Sommers. Since the ESS cross-section data were not available, this study adopts the 1.5 K thermal scattering cross section provided by PNPI for cold and thermal neutron transport calculations, focusing on the relationship between the average energy of emitted neutrons and the energy of incident neutrons.

\begin{figure}[!htbp]
    \centering

    \begin{minipage}[c]{0.48\linewidth}
        \centering
        \includegraphics[width=\linewidth]{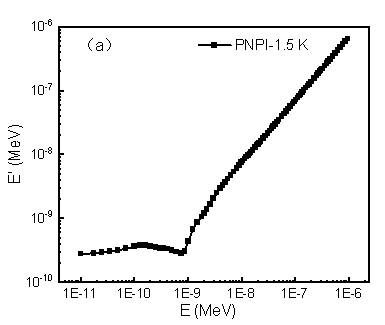}
    \end{minipage}
    \hfill
    \begin{minipage}[c]{0.48\linewidth}
        \centering
        \includegraphics[width=\linewidth]{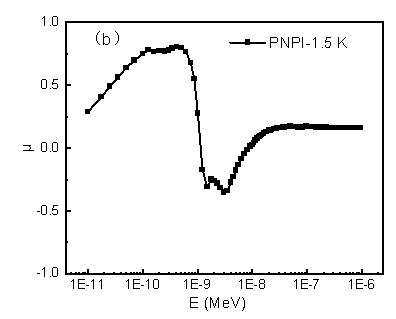}
    \end{minipage}

    \caption{Secondary neutron characteristics as a function of incident neutron energy in superfluid $^{4}\mathrm{He}$ at 1.5 K. Panel (a) shows the average energy, and panel (b) shows the average cosine of the emission angle \cite{bib:45}.}
    \label{CS_angle}
\end{figure}

As shown in Figure \ref{CS_angle} (a), the energy of emitted neutrons increases linearly once the incident neutron energy exceeds a certain threshold. Regarding the emission angle, as depicted in Figure \ref{CS_angle} (b), low-energy incident neutrons exhibit predominantly forward scattering with a tendency toward backward scattering, while high-energy neutrons result in nearly isotropic emission angles.

The nuclear cross-section libraries, scattering models, and temperatures used in the MCNPX simulations in this paper are summarized in Table \ref{Material}. The temperature-dependent thermal scattering data were generated using NJOY based on the corresponding scattering models.

\begin{table*}[htbp]
\centering
\caption{Material densities, nuclear cross-section libraries, scattering models, and temperatures used in the MCNPX simulations in this paper.}
\begin{tabular}{lcllc}
\toprule
Material &
\begin{tabular}{c}Density (g/cm$^3$)\end{tabular} &
Nuclear data (ZAID) &
Scattering model &
Temperature (K) \\
\midrule
Superfluid $^{4}\mathrm{He}$ & 0.145 & 2004.66c & $S(\alpha,\beta)$ & 0.5, 1.0, 1.5 \\
BeO & 2.8 & 4009.66c, 8016.66c & $S(\alpha,\beta)$ & 4, 7, 10, 15, 20 \\
Be & 1.85 & 4009.66c & $S(\alpha,\beta)$ & 4, 7, 10, 15, 20 \\
Al & 2.7 & 13027.64c & $S(\alpha,\beta)$ & 4, 7, 10, 15, 20 \\
CH$_4$ & 0.511 & 6000.66c, 1001.66c & $S(\alpha,\beta)$ & 4, 7, 10 \\
Solid D$_2$O & 1.099987 &  1002.66c, 8016.66c& Free-gas & 4, 7, 10, 15, 20\\
Ortho-D$_2$ & 0.1735 & 1002.66c & $S(\alpha,\beta)$ & 4, 7, 10, 15, 20 \\
H$_2$O & 0.995 & 1001.60c, 1002.60c, 8016.60c & Free-gas & 4, 7, 10, 15, 20 \\
Ortho-H$_2$ & 0.0725 & 1001.66c & $S(\alpha,\beta)$ & 4, 7, 10, 15, 20 \\
\bottomrule
\end{tabular}
\label{Material}
\end{table*}

We verified the simulation model against the prototype source at PIK \cite{bib:48}, obtaining good agreement in the simulated cold neutron distribution.

The thermal design and heat load analysis of the CSNS UCN source cryogenic system have been investigated separately \cite{bib:49}. Therefore, a detailed thermal analysis is beyond the scope of the present study, which focuses on the physical design and optimization of the UCN source.

\subsection{UCN production}\label{subsec:ucn-production}

\begin{figure}[!htb]
\centering
\includegraphics[width=\linewidth]{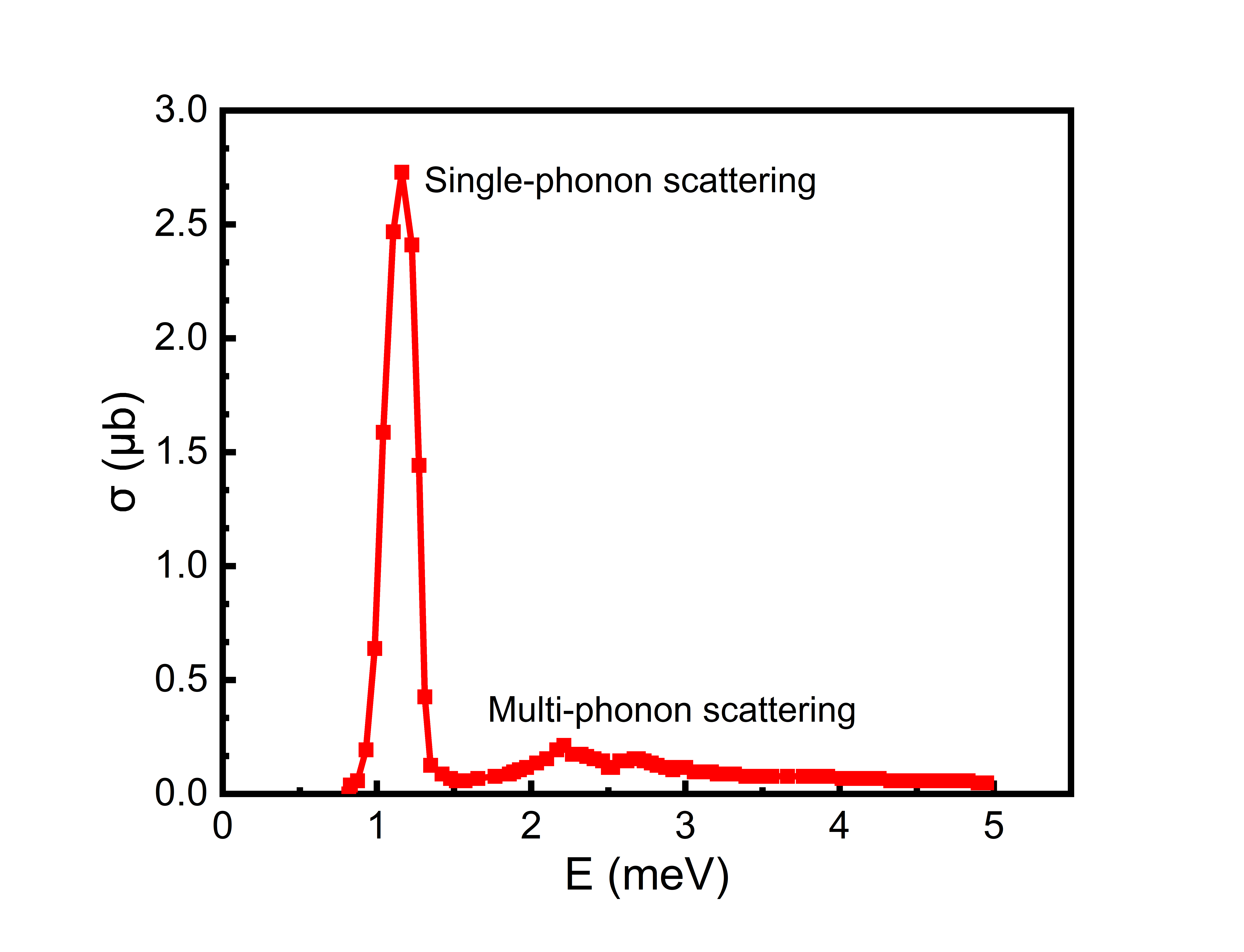}
\caption{Cross sections for the production of ultracold neutrons with energies up to 233.5 neV in superfluid $^{4}\mathrm{He}$ have been reported in the literature \cite{bib:14,bib:52}.}
\label{UCN_He_cs}
\end{figure}

The UCN production rate per unit volume \( P \) (\(\mathrm{UCN} \, \mathrm{cm}^{-3} \mathrm{s}^{-1}\)) is determined using a Python code based on the formula from Refs.~\cite{bib:14,bib:50}. Its expression is given by:
\begin{equation}
P = n_{^{4}\mathrm{He}} \int \varPhi(E) \, \sigma_{^{4}\mathrm{He}}(E) \, \mathrm{d}E
\label{P}
\end{equation}
where $n_{^{4}\mathrm{He}}$ ($\mathrm{cm}^{-3}$) represents the number density of $^{4}\mathrm{He}$. The term $\varPhi$ ($\mathrm{cm}^{-2}\mathrm{s}^{-1}\mathrm{meV}^{-1}$) signifies the incident cold neutron spectrum. $\sigma_{^{4}\mathrm{He}}$ ($\mathrm{cm}^{2}$) denotes the UCN production cross section in superfluid $^{4}\mathrm{He}$ for a maximum trapping energy of $E_{\mathrm{trap}} = V_{\mathrm{wall}} - V_{^{4}\mathrm{He}}$. For a Be-coated converter, the neutron optical potentials of Be and superfluid $^{4}\mathrm{He}$ are 252 neV and 18.5 neV, respectively, yielding $E_{\mathrm{trap}} = 233.5$ neV. The corresponding UCN production cross section is shown in Fig.~\ref{UCN_He_cs}. The cross section for producing UCNs from cold neutrons with energies below 0.8 meV is very low and can be neglected. Both single-phonon (8--10 \AA) and multiphonon (4--8 \AA) processes contribute to UCN production. UCN accumulation is slow because the production cross section is on the order of \si{\micro\barn} \cite{bib:14}. Because the available UCN phase space depends on the trapping energy $E_{\mathrm{trap}}$, the UCN production rate for other potentials can be estimated using the relationship $P \propto E_{\mathrm{trap}}^{3/2}$.
 
Alternatively, the single-phonon UCN production rate per unit volume can be expressed as \cite{bib:51,bib:52}:

\begin{equation}
P_{\mathrm{single}}
=\left.(1.39 \pm 0.11) \times 10^{-11}  E_{\mathrm{trap}}^{3 / 2} \frac{\mathrm{d}\phi}{\mathrm{d}\lambda_i}\right|_{\lambda^*}
\end{equation}

where $E_{\mathrm{trap}}$ is in units of $[\mathrm{neV}]$ and $\left.\frac{\mathrm{d}\phi}{\mathrm{d}\lambda_i}\right|_{\lambda^*}$ is the incoming intensity of cold neutrons at $\lambda^*=8.9 ~\text{\AA}$ in units of $\left[\mathrm{cm}^{-2}\,\mathrm{s}^{-1}\,\text{\AA}^{-1}\right]$.

We verified the simulation model against prototype sources at PIK \cite{bib:48}, SNS, and PULSTAR \cite{bib:52}, obtaining good agreement in calculated UCN yields.

\subsection{UCN storage and transport}\label{subsec:ucn-storage-transport}

The iMCUCN code is used to simulate the UCN storage and transport processes \cite{bib:36,bib:38}. We validated the simulation model against the existing SUN1 UCN source at ILL and observed a close alignment between the simulated and measured results. 

During the processes of UCN storage and transport, neutrons with kinetic energies lower than the optical potential of the wall material undergo total reflection upon colliding with the guide or converter walls. By coating these surfaces with materials characterized by a high optical potential, such as Ni, a quantum potential trap is established where UCNs can be effectively confined. Nevertheless, various loss mechanisms exist that limit the storage efficiency within this potential well. Incorporating these competing processes, the total storage time $\tau$ of UCNs in superfluid $^{4}\mathrm{He}$ is governed by the following relationship \cite{bib:18,bib:50}:

\begin{equation}
\tau^{-1}=\tau_{\mathrm{up}}^{-1}+\tau_{\mathrm{abs}}^{-1}+\tau_{\mathrm{wall}}^{-1}+\tau_{\beta}^{-1}+\tau_{\mathrm{leakage}}^{-1}+\tau_{\mathrm{impurities}}^{-1}
\label{ttotal}
\end{equation}

Here, the inverse of $\tau$ denotes the total loss rate. $\tau_{\mathrm{up}}^{-1}$ is affected by the UCN upscattering in superfluid $^{4}\mathrm{He}$ and is temperature-dependent. $\tau_{\mathrm{abs}}^{-1}$ originates from the significant neutron absorption cross section of $^{3}\mathrm{He}$ and is contingent on the abundance of $^{3}\mathrm{He}$.
When a UCN encounters a wall coated with a material of high Fermi potential, losses may occur due to wall absorption and upscattering ($\tau_{\mathrm{wall}}^{-1}$). The parameter $\tau_{\beta}$ = 880 s represents the neutron decay time. $\tau_{\mathrm{leakage}}^{-1}$ accounts for UCN leakage through gaps in the UCN shutter and converter walls. The corresponding loss rates can be expressed in terms of the macroscopic cross sections $\varSigma _i$ as follows:
\begin{equation}
\tau^{-1}=\sum_i{\varSigma _i}v 
\label{ttotalsigma}
\end{equation}

where $v$ is the UCN velocity. UCN flux can be expressed as $\phi_{\mathrm{UCN}}=v\rho$, where $\rho$ is the UCN density. The net UCN production rate per unit volume can be calculated as the UCN production rate determined by Eq. \eqref{P} minus the UCN loss associated with Eqs. \eqref{ttotal} or \eqref{ttotalsigma}. This relationship is expressed as :

\begin{equation}
{} P-P_{\mathrm{loss}}=P-\sum_i{\varSigma _i}\phi_{\mathrm{UCN}} =\frac{\mathrm{d}\rho}{\mathrm{d}t}
\end{equation}

\begin{figure}[h]%
\centering
\includegraphics[width=\linewidth]{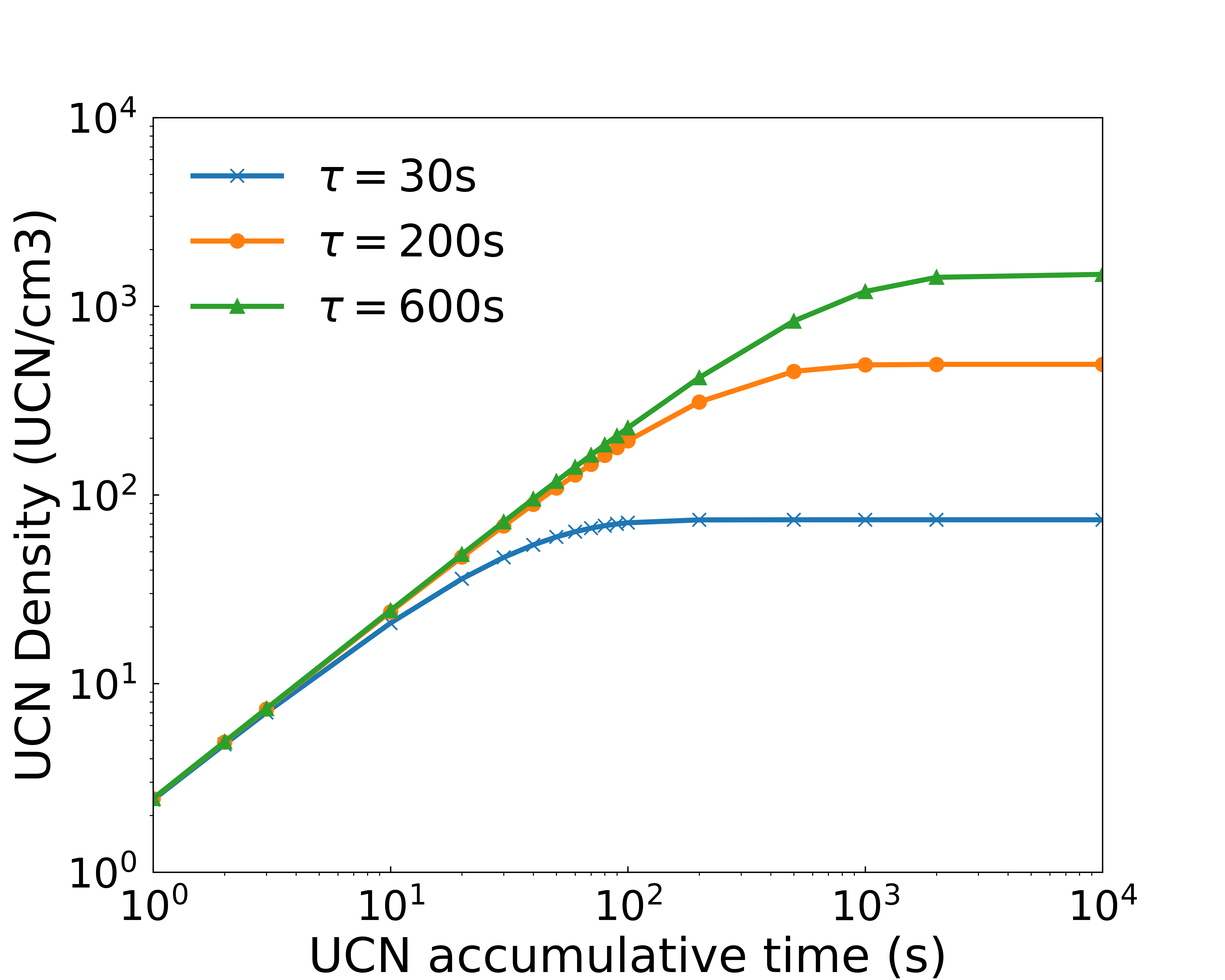}
\caption{UCN density as a function of accumulation time for different storage time constants.}
\label{UCN-density}
\end{figure}

The temporal evolution of the UCN density in a closed converter system is described by \cite{bib:50}:
\begin{equation}
{} \frac{\mathrm{d}\rho}{\mathrm{d}t}=P-\frac{\rho}{\tau} 
\end{equation}
Thus,
\begin{equation}
\rho =P\tau \left( 1-\mathrm{e}^{-\frac{t}{\tau}} \right)
\label{rho-t}
\end{equation}

Fig.~\ref{UCN-density} illustrates the UCN density as a function of the accumulation time. For a constant production rate and an initially empty converter, the UCN density reaches $31/32$ (approximately 96.9\%) of its steady-state value after $5\ln 2 \cdot \tau$. Once dynamic equilibrium is established, the saturated UCN density ($\rho$) within the converter is defined as \cite{bib:50}:

\begin{equation}
\rho =P\tau
\label{rho}
\end{equation}

The saturated UCN density is an important parameter for UCN experiments. For fixed converter and experimental-bottle volumes, a higher UCN density in the source leads to a higher UCN density and total neutron number in the experimental bottle, thereby reducing statistical uncertainties \cite{bib:50}. 

The UCN production rate, denoted by $R$ (UCN/s), is a crucial parameter for assessing a UCN source and is defined as:
\begin{equation}
R =PV
\label{R}
\end{equation}

When the UCN shutter is opened, let $N_{\mathrm{det}}$ denote the number of UCNs either detected by the detector or transmitted into the experimental bottle. The extraction process can then be simply quantified by the extraction efficiency $T$, defined as
\begin{equation}
    T = \frac{N_{\mathrm{det}}}{N_{\mathrm{tot}}}
\label{Trasmission}
\end{equation}
where $N_{\mathrm{tot}}$ is the total number of UCNs stored in the converter prior to extraction.

Following the accumulation-and-release concept described by Zimmer et al. for in-beam superfluid $^{4}\mathrm{He}$ UCN sources \cite{bib:53}, this preliminary study adopts the accumulation mode as the reference operating scheme for the UCN source. This operating mode is particularly suitable for UCN storage experiments. UCNs are accumulated in the converter with the UCN shutter closed until saturation, after which the incident cold neutron beam is switched off and the shutter is opened for extraction. Neglecting gravitational effects and the UCN energy distribution in this analytical approximation, the subsequent decay of the UCN population remaining in the converter is described by \cite{bib:36}:

\begin{equation}
N(t, E) = N_{\mathrm{sat}} \mathrm{e}^{-t(\frac{1}{\tau_{\mathrm{guide}}}+\frac{1}{\tau})}=P\tau V \mathrm{e}^{-t(\frac{1}{\tau_{\mathrm{guide}}}+\frac{1}{\tau})}
\label{nextraction}
\end{equation}

where $\tau_{\mathrm{guide}}$ is the guide leakage time constant after opening the shutter and is defined as \cite{bib:50}:
\begin{equation}  
\frac{1}{\tau_{\mathrm{guide}}} =\frac{S_{\mathrm{guide}}v}{4V} 
\label{eq:guide-extraction-time} 
\end{equation}

Here, $V$ represents the converter volume, $S_{\mathrm{guide}}$ is the area of the exit hole, and $v$ denotes the neutron velocity in vacuum. $\tau_{\mathrm{guide}}$ is dependent on losses in the guide and the energy of the UCNs, with higher-energy UCNs being more susceptible to losses compared to lower-energy ones \cite{bib:50}. According to Ref.~\cite{bib:36}, the UCN count rate reaching the detector, $C(t,E)$ (UCN/s), is proportional to the number of UCNs remaining in the converter, $N(t,E)$, divided by the extraction time constant, $\tau_{\mathrm{guide}}$, as expressed by

\[
C(t,E) \propto \frac{N(t,E)}{\tau_{\mathrm{guide}}}
\]

The proportionality accounts for the fact that not all UCNs leaving the converter reach the detector, as additional losses may occur during transport through the extraction guide.

The wall material parameters used in iMCUCN are summarized in Table \ref{material for iMCUCN}, based on published literature and personal communications \cite{bib:50,bib:54,bib:55,bib:56,bib:57,bib:58,bib:59}. The complex optical potential of a material can be expressed as \cite{bib:50}:

\begin{equation}  
U = V+\mathrm{i}W
\label{eq:complex-optical-potential} 
\end{equation}

Here, $V$ depends on the bound coherent scattering length and the number density of the material, while $W$ depends on losses due to absorption and inelastic scattering. When $W = 0$ and the perpendicular component of the UCN energy is lower than the wall optical potential, incident UCNs can undergo total reflection. When $W \neq 0$, total reflection no longer occurs; instead, the UCN beam undergoes a combination of partial reflection and concurrent loss due to wall interactions \cite{bib:50}. The ratio of $W$ to $V$ is related to the wall loss probability per bounce when UCNs collide with the material, thus defining the loss factor $\eta$ as \cite{bib:50}:
\begin{equation}  
\eta = \frac{W}{V}
\label{eq:wall-loss-factor} 
\end{equation}

\begin{table}[!htb]
\centering
\caption{Optical potentials, loss factors, and typical applications of UCN-related materials. G: guide; C: converter; R: reflector; B: storage bottle; DLC: diamond-like carbon.}
\setlength{\tabcolsep}{1.8pt}
\begin{tabular}{cccc}
\toprule
\multirow{4}{*}{Material} & 
\multirow{4}{*}{\begin{tabular}{c}Optical \\ potential \\$V$ ($\mathrm{neV}$)\end{tabular}} & 
\multirow{4}{*}{\begin{tabular}{c}UCN \\ loss factor \\ $\eta$ ($10^{-5}$)\end{tabular}} & 
\multirow{4}{*}{\begin{tabular}{c}Typical \\ applications \end{tabular}} \\
& & & \\
& & & \\
& & & \\
\midrule
Pure superfluid $^{4}\mathrm{He}$ & 18.5 & 0 & C \\
Replica & 220 & 14--25 & C \\
DLC & 220 & 0.0146 & C/G \\
BeO & 261 & 1.35 & C \\
Be & 252 & 0.5 & C/R \\
SS (polished stainless steel) & 183 & 9.3 & C/G \\
Fomblin ($\mathrm{C}_{3}\mathrm{F}_{6}\mathrm{O}$) & 115 & 1.85 & C \\
CYTOP ($\mathrm{C}_{6}\mathrm{F}_{10}\mathrm{O}$) & 115 & 2.7 & C \\
dPS (deuterated polystyrene) & 161 & 1.5--2 & C/B \\
$^{58}\mathrm{Ni}$ & 335 & 8.6 & G \\
$^{58}\mathrm{Ni}\mathrm{Mo}$ (15\% Mo) & 315 & 8.95 & G \\
$\mathrm{Ni}$ & 252 & 12.1 & G \\
$\mathrm{NiMo}$ & 220 & 12.0 & G \\
$\mathrm{NiP}$ (15\% P) & 213 & 11.0 & G \\
\bottomrule
\end{tabular}
\label{material for iMCUCN}
\end{table}

\section{Key factors affecting the design of a superfluid $^{4}\mathrm{He}$ UCN source}\label{sec:design-factors}

This section evaluates the principal design variables of the CSNS UCN source in three groups: the incident cold-neutron conditions, the converter parameters, and the extraction-system parameters. The objective is to identify a balanced configuration that provides a high saturated UCN density, total UCN production rate, and extraction efficiency, thereby establishing a preliminary physical design for the CSNS UCN source.

\begin{figure}[!htb]
\includegraphics
  [width=0.99\hsize]
  {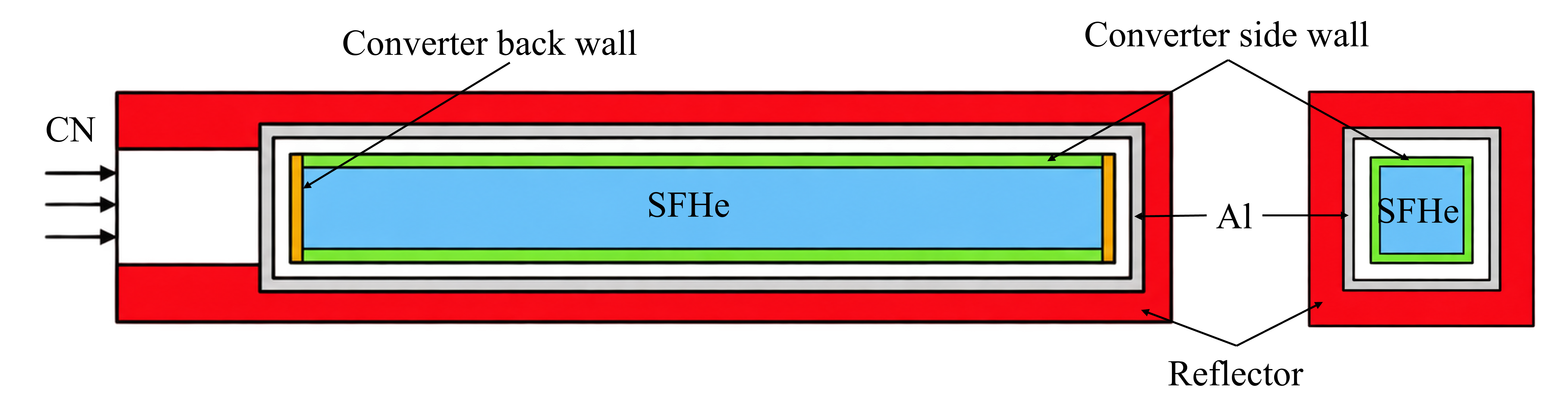}
\caption{The MCNP model of the “baseline” design  at Beamline 12 of CSNS.}
\label{MCNPXModel}
\end{figure}

\begin{figure}[!htb]
\includegraphics
  [width=0.9\hsize]
  {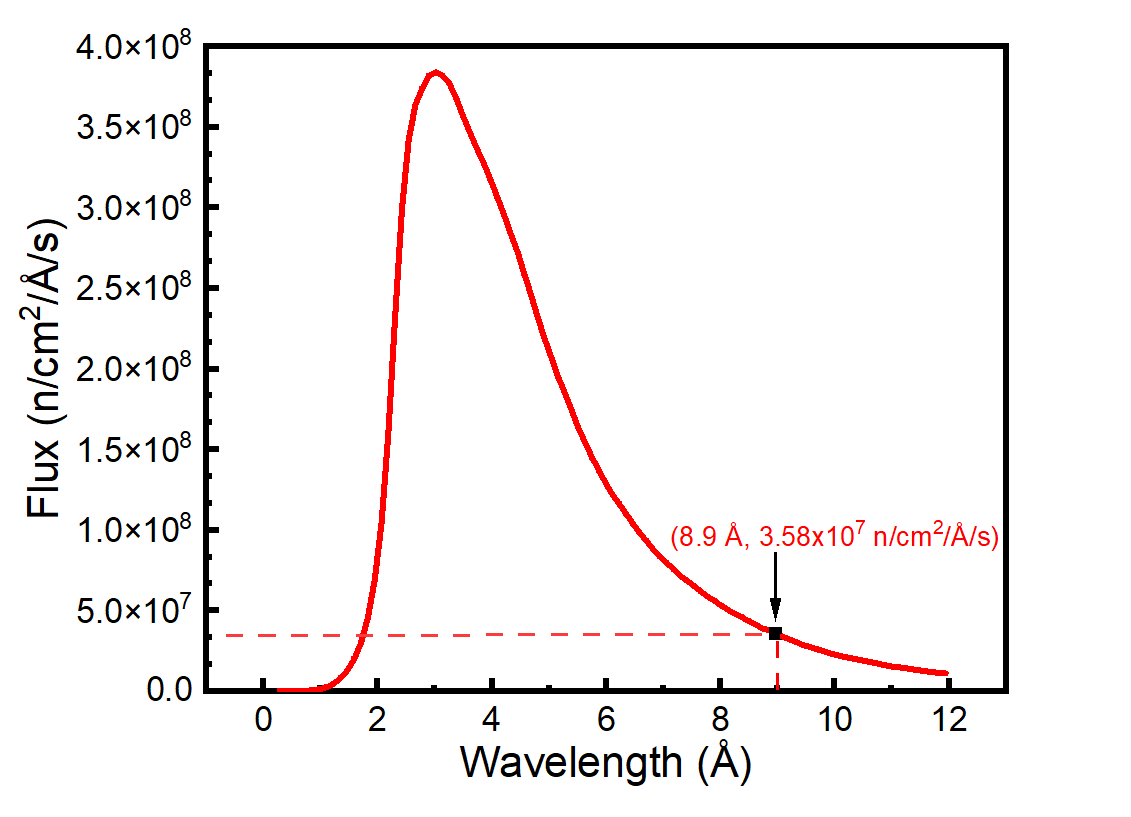}
\caption{The incident cold neutron spectrum at Beamline 12 of CSNS.}
\label{fig:IncidentCN}
\end{figure}

For the cold-neutron transport calculations, the reference MCNPX geometry and incident wavelength spectrum are shown in Figs.~\ref{MCNPXModel} and \ref{fig:IncidentCN}, respectively. Except for the guide--converter cross-section scan, the reference model consists of a \SI{100}{\cm}-long superfluid $^{4}\mathrm{He}$ converter with a square cross section of $\SI{7}{\cm}\times\SI{7}{\cm}$. The converter is enclosed by \SI{0.50}{\cm}-thick BeO side walls, \SI{0.10}{\cm}-thick Be end windows, a \SI{1.00}{\cm}-thick intermediate region, and a \SI{0.15}{\cm}-thick Al vessel. The outer region is reserved for the moderator--reflector studies and forms a \SI{20.60}{\cm}-long upstream entrance section containing a $\SI{7}{\cm}\times\SI{7}{\cm}$ cold-neutron channel. The wavelength spectrum is applied at the upstream entrance of this channel and corresponds to an integrated cold-neutron intensity of $6.71\times10^{10}~\mathrm{n/s}$ over the wavelength range of 0.10--12.00~\text{\AA}.

For the initial parameter studies, an effective UCN storage time constant of $\tau=200~\mathrm{s}$ was assumed to isolate the effects of cold-neutron transport and converter geometry. The cold-neutron spectrum within the superfluid $^{4}\mathrm{He}$ converter was calculated using MCNPX, and the UCN production rate per unit volume $P$, saturated UCN density $\rho_{\mathrm{sat}}$, and total UCN production rate $R$ were determined using Eqs.~\eqref{P}, \eqref{rho}, and \eqref{R}, respectively. Unless otherwise specified, Be was used as the reference converter wall material, and the corresponding UCN production cross section is shown in Fig.~\ref{UCN_He_cs}. At this stage, the UCN extraction system was not included, and the individual UCN loss mechanisms were not evaluated separately. These effects are subsequently incorporated into the iMCUCN simulations, with the storage time constant recalculated for different converter wall materials, as described in Section~\ref{subsubsec:converter-wall-material}.

The source design was refined sequentially. At each stage, the parameter or parameter group under investigation was evaluated relative to the current reference configuration, and the resulting design decision was carried forward to define the reference configuration for the subsequent stage.

\subsection{Incident cold neutrons}\label{subsec:incident-cold-neutrons}

This section focuses on the optimization of the UCN source by analyzing the contributions of cold neutrons at different wavelengths,
the incident cold neutron guide cross section, and the effects of the moderator-reflector system. 
For in-beam UCN sources, the total cold neutron intensity entering the converter (n/s) is a more relevant figure of merit for UCN production than local neutron brightness (n/cm$^2$/s) \cite{bib:31,bib:32}. According to Eqs. \eqref{P}, \eqref{rho} and \eqref{R}, a higher cold neutron intensity directly leads to a greater saturated UCN density and a higher UCN production rate within the converter.

Since the UCN accumulation time constant in superfluid $^{4}\mathrm{He}$ can reach tens to hundreds of seconds, much longer than the pulse intervals of tens of milliseconds, the distinction between continuous and pulsed neutron sources has a negligible impact on UCN accumulation. Therefore, the UCN production rate at CSNS is calculated using the time-averaged incident cold neutron flux. Furthermore, applying a cold neutron reflection layer to the converter surface renders the incident angle of the cold neutrons negligible in terms of its impact on production efficiency. Consequently, the incident angle will not be discussed in detail.

\subsubsection{Contribution of cold neutrons with different wavelengths to ultracold neutron production}\label{subsubsec:wavelength-contributions}

In-beam UCN sources such as SUN1 \cite{bib:18,bib:19} and SUN2 \cite{bib:20} utilize a monochromator to select cold neutrons near 8.9 \text{\AA} as the primary source, whereas SuperSUN \cite{bib:21} and PIK \cite{bib:48} utilize the full cold neutron spectrum. Ref.~\cite{bib:18} reports that the differential neutron flux at 8.9~\text{\AA} in the direct white beam at the exit of guide H172 is approximately five times higher than that in the monochromated secondary beam.

UCNs are primarily generated via single-phonon processes from cold neutrons near 8.9 \text{\AA}, as well as through multiphonon processes driven by neutrons in the 4--8 \text{\AA} range, as illustrated in Fig. \ref{UCN_He_cs}. Using the incident cold neutron wavelength spectrum at the Neutron Physics and Application Spectrometer at CSNS, shown in Fig. \ref{fig:IncidentCN}, the relative contributions of these processes were quantified. Single-phonon processes account for approximately 66\% of the total UCN production rate, while multiphonon processes contribute the remaining 34\%, as summarized in Table \ref{Wavelength}.  Given that the latter represents a significant portion of the output and considering the flux losses associated with monochromatization, the CSNS UCN source adopts the full-spectrum approach to maximize the overall UCN production rate.

\begin{table}[!htb]
\centering
\caption{Contributions of cold neutrons at different wavelengths to UCN production at CSNS Beamline 12.}
\begin{tabular}{cc}
\toprule
Process & Relative contribution to UCN production \\
\midrule
Single-phonon & $65.9\%$  \\
Multiphonon & $34.1\%$ \\
\bottomrule
\end{tabular}
\label{Wavelength}
\end{table}

\subsubsection{Incident cold neutron guide cross section}\label{subsubsec:incident-guide-cross-section}

The cross-sectional area of the incident cold neutron guide influences the total cold neutron intensity entering the converter. To investigate its effect on UCN production, the guide and converter were assigned identical square cross sections, with their dimensions varied simultaneously while maintaining a fixed converter length of 1 m. The dependence of the storage time constant on the converter geometry was therefore not considered in this comparison.

Table \ref{spotsize} presents the saturated UCN density and total UCN production rate as functions of the incident guide cross section. As the cross-sectional area increases, the total incident cold neutron intensity and UCN production rate increase, whereas the saturated UCN density decreases due to the corresponding increase in converter volume. According to Ref. \cite{bib:55}, the design of a UCN source should balance a high saturated UCN density, which benefits experimental counting statistics, with a sufficiently large converter volume to support efficient UCN extraction.

Considering these competing requirements, a guide cross section of $7 \times 7$ cm$^2$ was selected, as the UCN production rate approaches a plateau while maintaining a relatively high saturated UCN density. This choice is consistent with the converter dimensions adopted in previous UCN source designs, including those of Golub \cite{bib:60}, SUN1 \cite{bib:18,bib:19}, SUN2 \cite{bib:20}, and SuperSUN \cite{bib:21} at the ILL. Further improvements in saturated UCN density may be achieved by increasing the UCN storage time constant.

\begin{table*}[!htb]
\centering
\caption{Saturated UCN density and total UCN production rate as functions of the incident cold neutron guide and converter cross-sectional dimensions, assuming a constant UCN storage time constant of $\tau = 200\,\mathrm{s}$ for all configurations.}

\setlength{\tabcolsep}{4pt}
\renewcommand{\arraystretch}{1.0}

\begin{tabular}{ccccc}
\toprule
\begin{tabular}{c}
Converter cross-sectional\\
dimensions (cm $\times$ cm)
\end{tabular}
&
\begin{tabular}{c}
Cold neutron guide\\
cross-sectional dimensions\\
(cm $\times$ cm)
\end{tabular}
&
\begin{tabular}{c}
Incident cold neutron\\
intensity (n/s)
\end{tabular}
&
\begin{tabular}{c}
Saturated UCN density\\
(UCN/cm$^{3}$)
\end{tabular}
&
\begin{tabular}{c}
UCN production rate\\
(UCN/s)
\end{tabular}
\\
\midrule
$2 \times 2$   & $2 \times 2$   & $2.80 \times 10^{10}$ & 2915 & 5831 \\
$5 \times 5$   & $5 \times 5$   & $5.83 \times 10^{10}$ & 886  & 11078 \\
$7 \times 7$   & $7 \times 7$   & $6.71 \times 10^{10}$ & 492  & 12055 \\
$10 \times 10$ & $10 \times 10$ & $7.21 \times 10^{10}$ & 254  & 12678 \\
$12 \times 12$ & $12 \times 12$ & $7.21 \times 10^{10}$ & 178  & 12809 \\
\bottomrule
\end{tabular}
\label{spotsize}
\end{table*}

\subsubsection{Incident cold-neutron angular divergence}\label{subsubsec:incident-divergence}

The angular divergence of incident cold neutrons affects their trajectories through the converter and, consequently, the probability of leaving the conversion volume before undergoing UCN-producing scattering. To examine this effect, the incident beam cross section was fixed at $7\times7~\mathrm{cm^2}$, and three square converter cross sections ($7\times7$, $8\times8$, and $10\times10~\mathrm{cm^2}$) were compared for incident angular divergences between $0^\circ$ and $3^\circ$. The converter length was fixed at $1~\mathrm{m}$, and the preliminary effective storage time constant was set to $\tau=200~\mathrm{s}$.

\begin{figure*}[htbp]
    \centering
    \subfloat[Saturated UCN density.]{
        \includegraphics[width=0.45\textwidth]{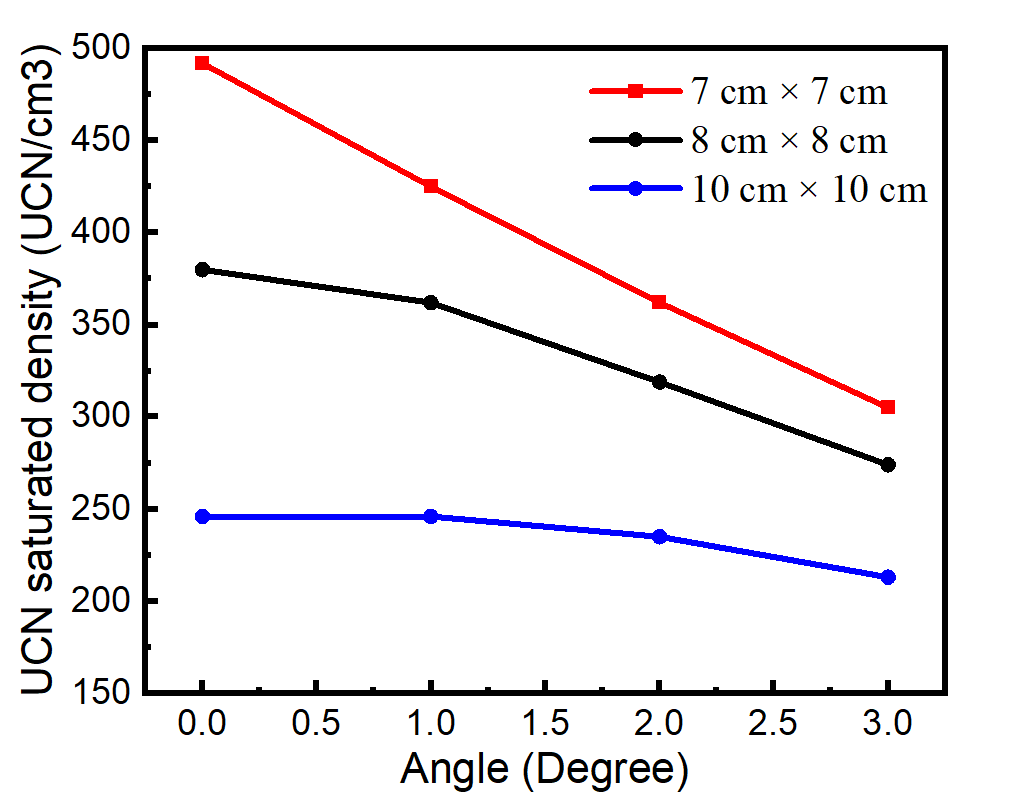}
    }
    \hfill
    \subfloat[Total UCN production rate.]{
        \includegraphics[width=0.45\textwidth]{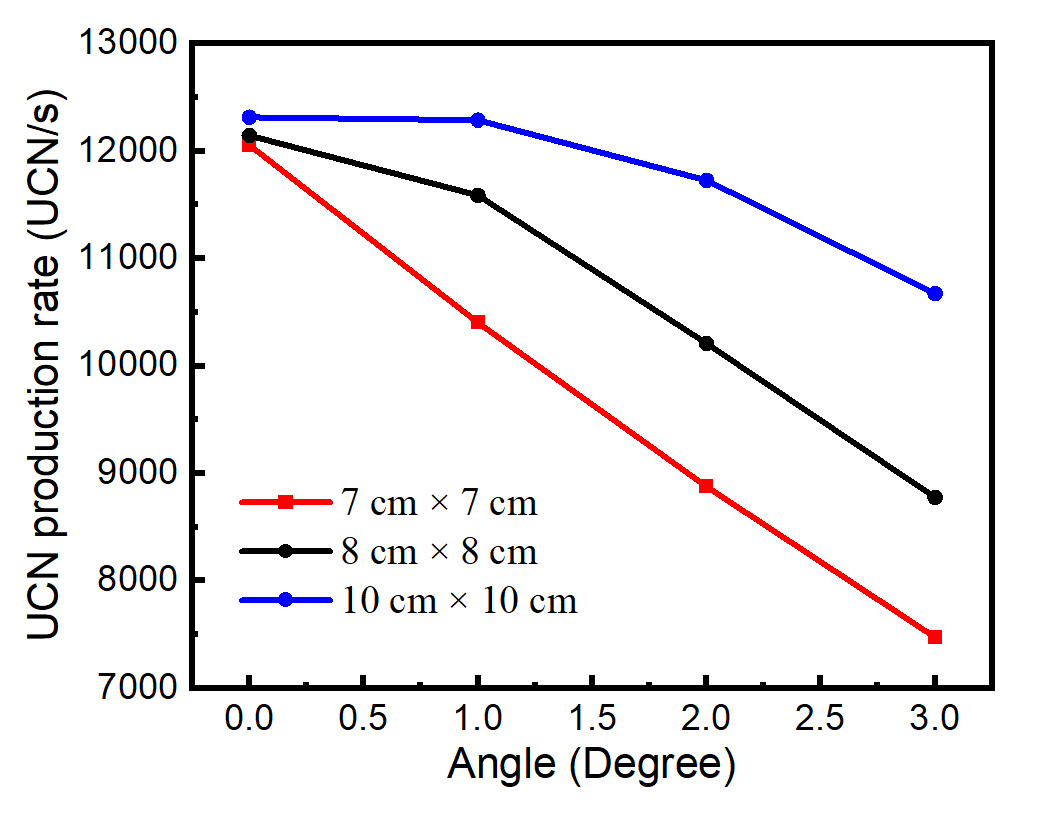}
    }
    \caption{Effect of incident cold-neutron angular divergence on (a) saturated UCN density and (b) total UCN production rate for converter cross sections of $7\times7$, $8\times8$, and $10\times10~\mathrm{cm^2}$. The incident cold-neutron beam cross section is fixed at $7\times7~\mathrm{cm^2}$, with a $1~\mathrm{m}$ converter length and an assumed effective storage time constant of $\tau=200~\mathrm{s}$.}
    \label{fig:incident-divergence}
\end{figure*}

As shown in Fig.~\ref{fig:incident-divergence}, both the saturated UCN density and the total UCN production rate generally decrease as the incident angular divergence increases. Increasing the converter cross section reduces the sensitivity of the total production rate to beam divergence, but also lowers the saturated UCN density.

Previous superfluid $^{4}\mathrm{He}$ UCN sources at ILL, including SUN1 \cite{bib:18,bib:19}, SUN2 \cite{bib:20}, and SuperSUN \cite{bib:21,bib:61}, have employed neutron-reflecting converter walls to increase the path length of cold neutrons within the converter and thereby enhance UCN production. Following this approach, Be is considered as a candidate converter wall material for the CSNS UCN source to enhance cold-neutron reflection and UCN production. Although the incident angular divergence of the CSNS Beamline 12 cold-neutron beam is estimated to be approximately $1^\circ$, an idealized parallel incident beam ($0^\circ$ divergence) was assumed in the subsequent preliminary calculations. The divergence-dependent reduction shown in Fig.~\ref{fig:incident-divergence} is therefore not included in those predictions.

\subsubsection{Moderator-reflector}\label{subsubsec:moderator-reflector}

The concept of enhancing a superthermal UCN source by surrounding the converter with a reflector cavity of high cold-neutron albedo was proposed by Bangert et al. \cite{bib:62}. In the PIK configuration, thermal neutrons from the reactor are moderated into cold neutrons within the moderator-reflector system and subsequently reflected into the UCN converter to enhance UCN production \cite{bib:48,bib:63}.

In this subsection, the gain in saturated UCN density achieved by different moderator-reflector materials at various temperatures and thicknesses is investigated. The selected full-spectrum cold-neutron configuration and the
$7\times7~\mathrm{cm^2}$ guide--converter cross section were
retained, with an assumed UCN storage time constant of
$\tau=200~\mathrm{s}$. The materials, nuclear cross-section data, scattering models, and corresponding temperatures used in the MCNPX simulations are summarized in Table \ref{Material}. Since phonon frequency spectra for many materials are unavailable at low temperatures, the cross sections of solid $\mathrm{D}_{2}\mathrm{O}$ and $\mathrm{H}_{2}\mathrm{O}$ were calculated using NJOY based on the free-gas model. Solid $\mathrm{D}_{2}\mathrm{O}$ was also employed as a cold moderator surrounding the superfluid $^{4}\mathrm{He}$ converter in the RCNP UCN source design \cite{bib:64}. In addition, the cross sections of ortho-$\mathrm{D}_{2}$ and ortho-$\mathrm{H}_{2}$ were calculated using their phonon spectra at 20 K \cite{bib:65}. The cross section of solid methane at 4 K was calculated using its phonon spectrum and has been experimentally validated \cite{bib:66,bib:67}.

The optimal temperatures and thicknesses, along with the corresponding gains in saturated UCN density, are summarized in Table \ref{moderator_reflector}. Solid $\mathrm{D}_{2}\mathrm{O}$ with a thickness of 150 cm at 7 K achieves the highest gain in saturated UCN density of 26 \%. This is followed by 150 cm of solid deuterium at 10 K with a gain of 19 \%, and 3 cm of solid methane at 4 K with a gain of 18 \%.

\begin{table}[!htb]
\centering
\caption{Optimized temperatures and thicknesses of candidate moderator-reflector materials and the corresponding gains in saturated UCN density, assuming a constant UCN storage time constant of $\tau = 200\,\mathrm{s}$ for all configurations.}
\begin{tabular}{cccc}
\toprule
\multirow{4}{*}{Material} & 
\multirow{4}{*}{\begin{tabular}{c}Temperature \\ (K)\end{tabular}} & 
\multirow{4}{*}{\begin{tabular}{c}Thickness \\ (cm)\end{tabular}} & 
\multirow{4}{*}{\begin{tabular}{c}Saturated UCN density \\ gain (\%)\end{tabular}} \\
& & & \\
& & & \\

\midrule
Solid $\mathrm{CH}_{4}$ & 4 & 3 & 18 \\
Solid $\mathrm{D}_{2}\mathrm{O}$ & 7 & 150 & 26 \\
Ortho-$\mathrm{D}_{2}$ & 10 & 150 & 19 \\
$\mathrm{H}_{2}\mathrm{O}$ & 10 & 3 & 9 \\
Ortho-$\mathrm{H}_{2}$ & 4 & 5 & 10 \\
\bottomrule
\end{tabular}
\label{moderator_reflector}
\end{table}

In contrast to PIK, where the incident thermal neutrons must be moderated into the cold neutron wavelength range required for efficient UCN production, the CSNS UCN source is supplied directly with cold neutrons from a 20 K coupled liquid hydrogen moderator. Consequently, the moderator-reflector system plays a more critical role at PIK, whereas its additional contribution at CSNS is relatively limited. Considering the enhancement in UCN density obtained in the present simulations and the associated engineering complexity, the moderator-reflector system is omitted from the initial CSNS UCN source design.

\subsection{The converter}\label{subsec:converter}

UCNs are generated and stored within the converter system. According to Eqs. \eqref{ttotal}, \eqref{rho}, and \eqref{R}, the parameters of the converter are critical to determining both the UCN loss probability and the total storage time constant, thereby directly influencing the saturated UCN density and the number of stored UCNs. The efficiency of the converter system is primarily governed by its geometric configuration (shape and length), the temperature of the superfluid $^{4}\mathrm{He}$, the abundance of $^{3}\mathrm{He}$ impurities within the superfluid $^{4}\mathrm{He}$, the leakage through gap areas, and the choice of wall materials. For the purposes of this analysis, the potential impact of surface impurities on the converter walls is disregarded.

\subsubsection{Converter shape}\label{subsubsec:converter-shape}

Rectangular converter geometries have been used in SUN1 and SUN2 \cite{bib:18,bib:19,bib:20}, while a cylindrical geometry is adopted in SuperSUN \cite{bib:21,bib:61}. Spherical converter geometries have also been considered in UCN source studies related to PIK \cite{bib:48,bib:63}. To compare their performance for the CSNS source, converter geometries with the same superfluid $^{4}\mathrm{He}$ volume of $4.9~\mathrm{L}$ were evaluated. The incident cold-neutron wavelength spectrum and the $7\times7~\mathrm{cm^2}$ guide cross section were kept unchanged, while only the converter shape was varied. An effective UCN storage time constant of $\tau=200~\mathrm{s}$ was assumed for all configurations. The different converter geometries also permit different arrangements of the extraction guide; their UCN production performance is compared here without incorporating the extraction efficiency into the reported figures of merit.

The calculated results are summarized in Table~\ref{tab:converter-shapes}. The rectangular converter provides the highest saturated UCN density and total production rate among the cases considered, whereas the spherical converter gives the lowest values, consistent with the shorter neutron paths through its conversion volume. Accordingly, the rectangular geometry was retained for the subsequent converter optimization.

\begin{table*}[htbp]
\centering
\caption{Comparison of UCN converter geometries with a fixed superfluid $^{4}\mathrm{He}$ volume of 4.9~L and an assumed effective UCN storage time constant of $\tau=200~\mathrm{s}$.}
\label{tab:converter-shapes}
\begin{tabular}{lcc}
\toprule
Converter geometry &
Saturated UCN density (UCN/cm$^3$) &
UCN production rate (UCN/s) \\
\midrule
Rectangular ($7\times7\times100~\mathrm{cm^3}$) & 492 & 12055 \\
Cylindrical ($r=3.949~\mathrm{cm}$, $L=100~\mathrm{cm}$) & 446 & 10974 \\
Cylindrical ($r=4.95~\mathrm{cm}$, $L=63.6~\mathrm{cm}$) & 333 & 8127 \\
Spherical ($r=10.54~\mathrm{cm}$) & 117 & 2833 \\
\bottomrule
\end{tabular}
\end{table*}

\subsubsection{The converter length}\label{subsubsec:converter-length}

\begin{figure}[!htb] \centering \includegraphics[width=\linewidth]{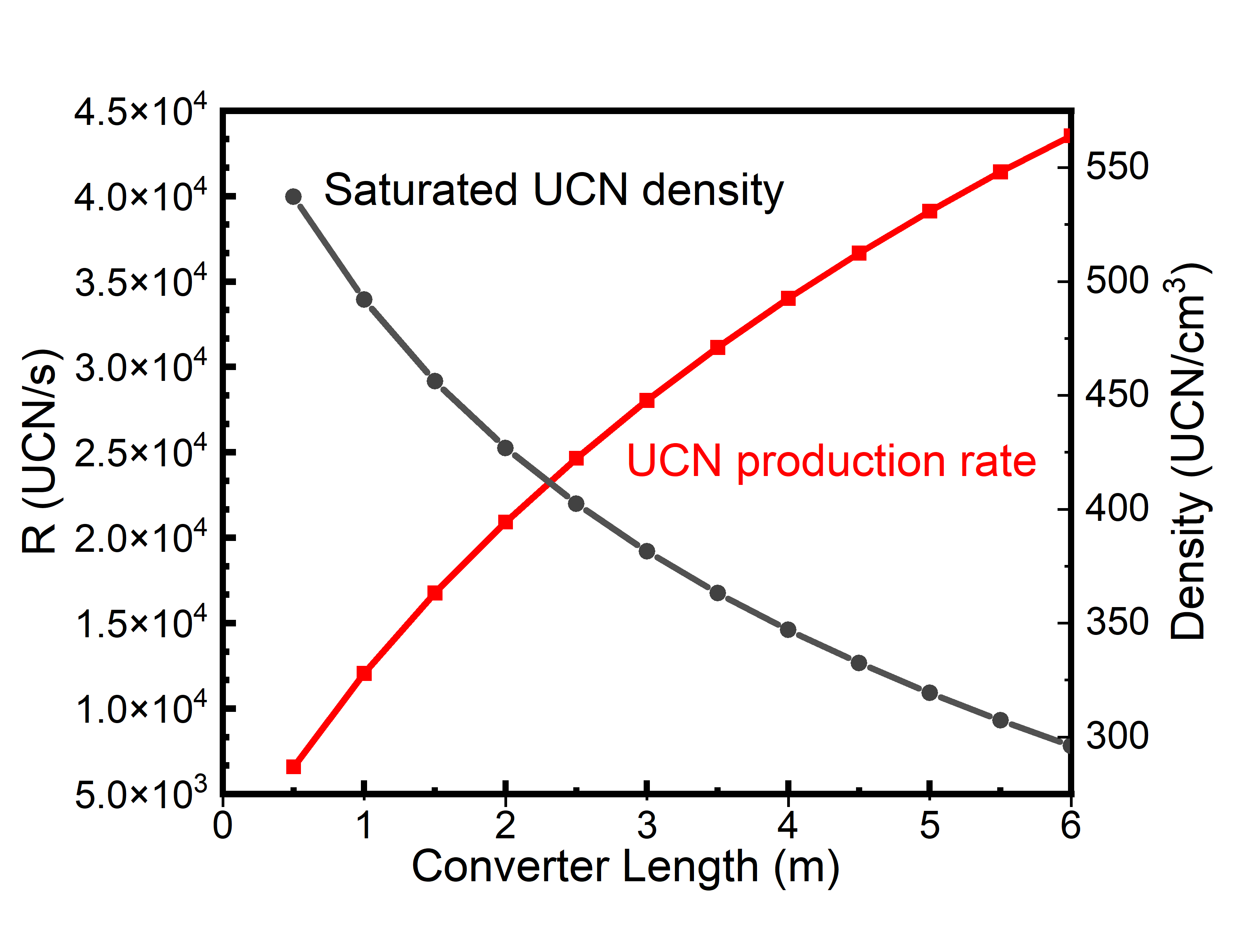} \caption{Functional dependence of the saturated UCN density and UCN production rate on the converter length, assuming a constant UCN storage time constant of $\tau = 200\,\mathrm{s}$ for all configurations.}\label{converter_length} \end{figure}

As illustrated in Fig. \ref{converter_length}, an increase in converter length results in a higher UCN production rate. The selected full-spectrum cold-neutron configuration and the $7\times7~\mathrm{cm^2}$ converter cross section were retained, with an assumed UCN storage time constant of $\tau=200~\mathrm{s}$. However, the saturated UCN density decreases as the converter length increases from 0.5~m to 6~m. This phenomenon occurs because the contribution from single-phonon processes remains relatively stable, owing to the long mean free path of 8.9~\AA\ cold neutrons (approximately 17~m in superfluid helium \cite{bib:44}). In contrast, the contribution from multiphonon processes diminishes with increasing converter length, due to the shorter mean free path of cold neutrons in the 4--8~\AA\ range.

Although a longer converter can achieve a higher UCN production rate, a converter length of 1~m was selected for the initial design, considering the spatial constraints and engineering requirements at CSNS.

\subsubsection{The superfluid $^{4}\mathrm{He}$ temperature}\label{subsubsec:helium-temperature}

In the preceding preliminary parameter studies, a constant effective UCN storage time constant of $\tau=200~\mathrm{s}$ was assumed. However, as described by Eq.~\eqref{ttotal}, the total storage time constant depends on several independent UCN loss mechanisms. In the following subsections, the contributions from upscattering, $^{3}\mathrm{He}$ absorption, leakage, and wall losses are examined individually through analytical or numerical calculations. These loss mechanisms are subsequently incorporated into the iMCUCN simulations to evaluate the total storage time constant for different converter wall materials.

UCNs can be upscattered into cold neutrons through excitations within the superfluid $^{4}\mathrm{He}$. The total upscattering loss rate $\tau_{\mathrm{up}}^{-1}$ comprises contributions from  one-phonon absorption, two-phonon scattering (involving absorption and emission of phonons), and roton-phonon scattering \cite{bib:68}. According to H. Yoshiki \cite{bib:60, bib:69}, the upscattering loss rate as a function of temperature $T$ is expressed as:

\begin{equation}
\frac{1}{\tau_{\mathrm{up}}}=A \mathrm{e}^{\frac{-12} {T} }+B T^{7}+C T^{\frac{3} {2} } \mathrm{e}^{-\frac{8.6}{T} } \,,
\label{upscattering-temperature}
\end{equation}

where the first term originates from one-phonon absorption, the second from two-phonon scattering, and the third from roton-phonon scattering.  The second term makes the dominant contribution to the upscattering loss rate (the inverse of $\tau_{\mathrm{up}}$).

\begin{figure}[!htb]\centering\includegraphics[width=\linewidth]{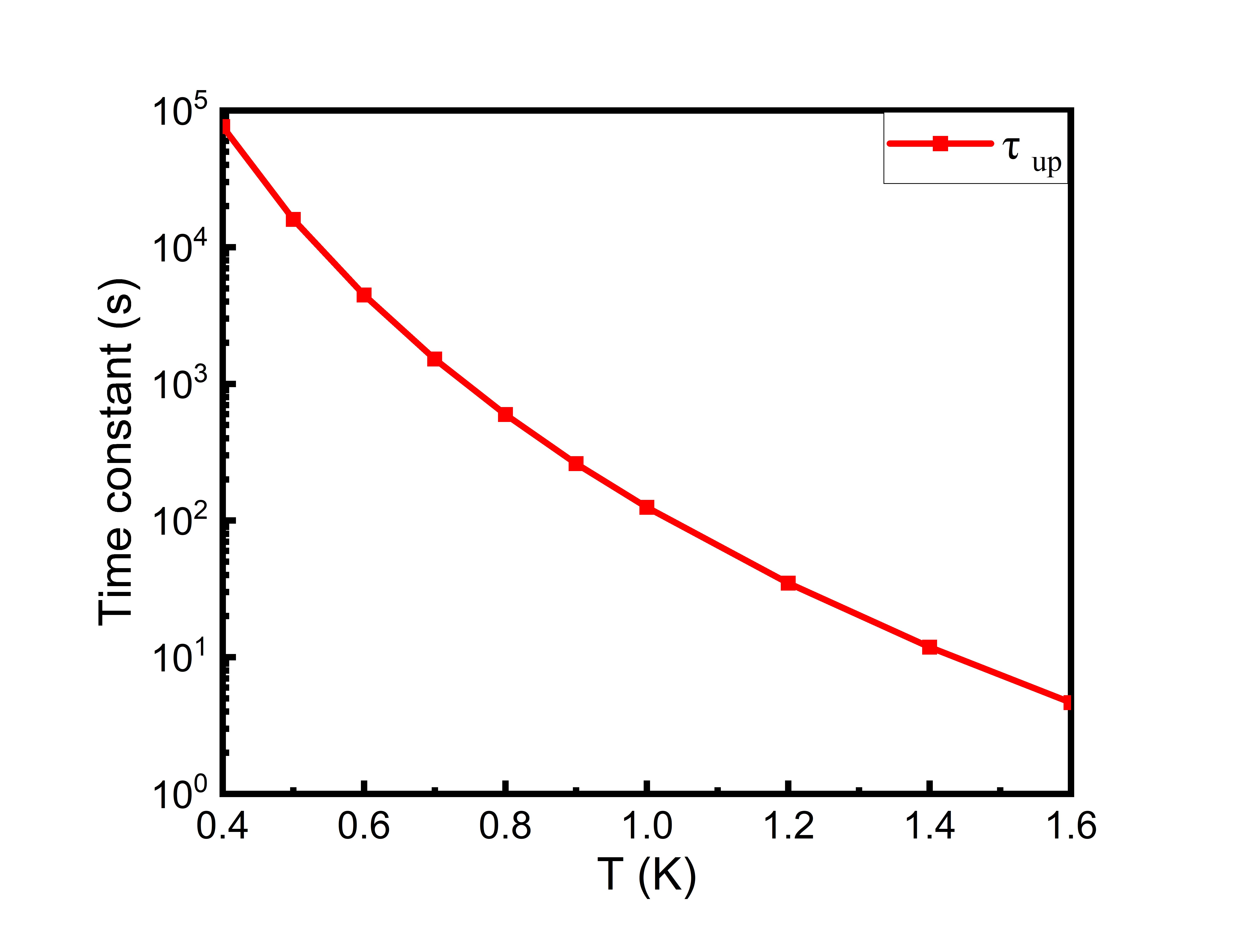}\caption{The upscattering time constant $\tau_{\mathrm{up}}$ as a function of the superfluid $^{4}\mathrm{He}$ temperature $T$.}\label{temperature}\end{figure}

The temperature dependence of the upscattering time constant $\tau_{\mathrm{up}}$, calculated directly from Eq.~\ref{upscattering-temperature}, is shown in Fig.~\ref{temperature}. At $T = 0.7$~K, with coefficients $A=130~\mathrm{s}^{-1}$ \cite{bib:68,bib:70}, $B=9.0 \times 10^{-3}~\mathrm{s}^{-1} \mathrm{K}^{-7}$, and $C=18~\mathrm{s}^{-1} \mathrm{K}^{-3/2}$, the resulting time constant is $\tau_{\mathrm{up}} = 1247$~s. In contrast, at $T = 2$~K, the time constant drastically decreases to $\tau_{\mathrm{up}} = 0.36$~s. Since upscattering contributes to the total UCN loss rate, maintaining a low temperature is essential to maximize the saturated UCN density.

\subsubsection{The $ ^{3}\mathrm{He}$ in superfluid $ ^{4}\mathrm{He}$}\label{subsubsec:helium3-impurity}

The natural abundance ratio of $n_{{}^{3}\mathrm{He}}$ to $n_{{}^{4}\mathrm{He}}$ is approximately $1.37 \times 10^{-6}$ and the actual $^{3}\mathrm{He}$ concentration in $^{4}\mathrm{He}$ may vary depending on the source of the $^{3}\mathrm{He}$. 
In superfluid $^{4}\mathrm{He}$, $^{3}\mathrm{He}$ atoms act as potent neutron absorbers, thereby reducing the storage time constant within the converter. 
The absorption time constant $\tau_{\mathrm{abs}}$ is given by:
\begin{equation}  
\frac{1}{\tau_{\mathrm{abs}}} 
= n_{{}^{3}\mathrm{He}} \, \sigma_{\mathrm{a}} \, v 
= n_{{}^{3}\mathrm{He}} \, \sigma_{\mathrm{a(th)}} \, v_{\mathrm{(th)}} ,
\label{tHe3}  
\end{equation}
where $n_{{}^{3}\mathrm{He}}$ denotes the number density of $^{3}\mathrm{He}$ atoms. Due to the $1/v$ dependence of the neutron absorption cross section, this expression remains independent of the UCN velocity.
For $^{3}\mathrm{He}$, the absorption cross section $\sigma_{\mathrm{a}}$ is $5333~\mathrm{b}$ at a neutron velocity of $2200~\mathrm{m/s}$. 

\begin{figure}[!htb]
\centering
\includegraphics[width=\linewidth]{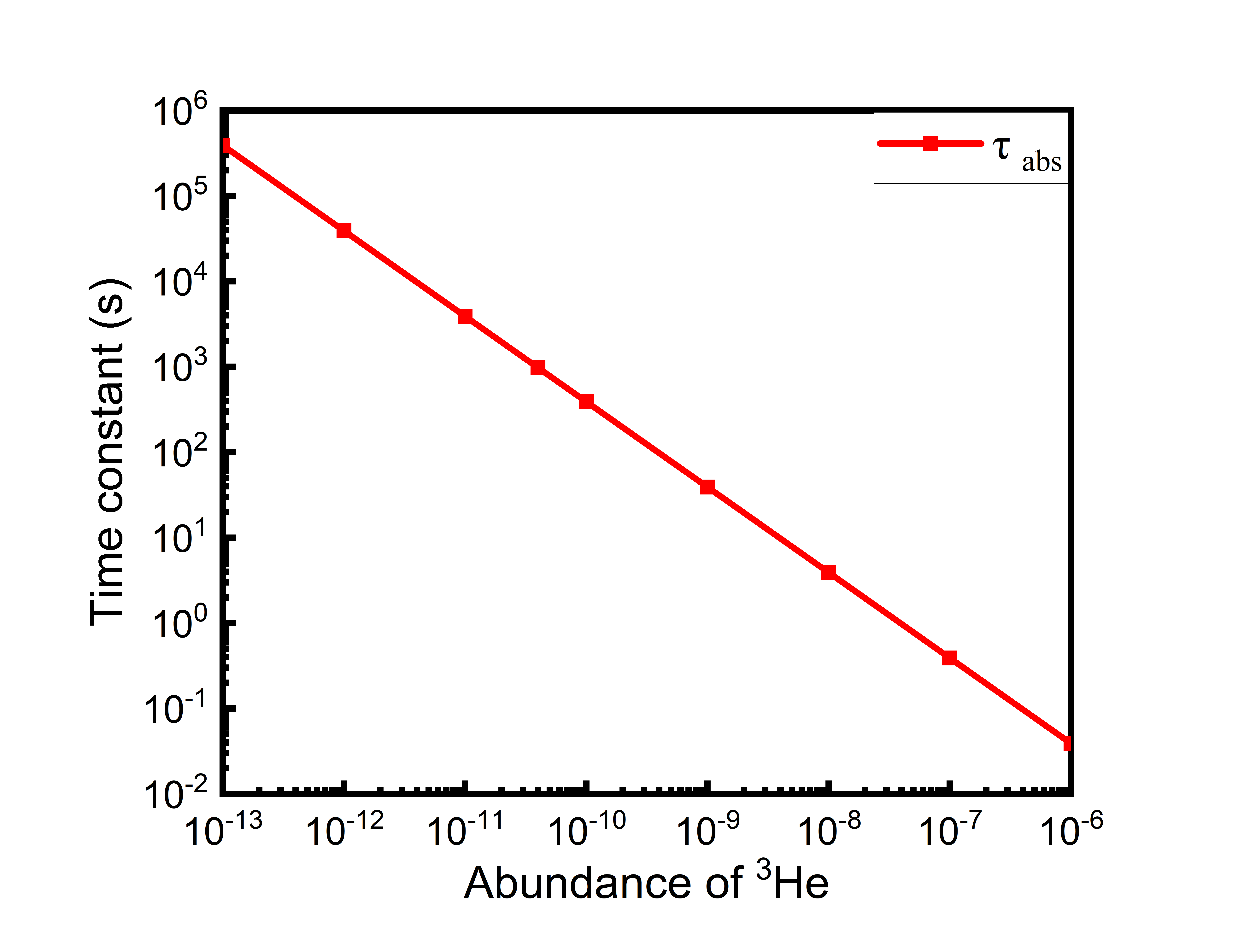}
\caption{The absorption time constant $\tau_{\mathrm{abs}}$ as a function of the abundance of $^{3}\mathrm{He}$.}\label{abun_he3}
\end{figure}

The dependence of the absorption time constant $\tau_{\mathrm{abs}}$ on the $^{3}\mathrm{He}$ abundance, calculated directly from Eq.~\eqref{tHe3}, is shown in Fig.~\ref{abun_he3}. The absorption time constant is independent of the UCN energy. To prolong the UCN storage duration and maximize the saturated UCN density, it is essential to minimize the $^{3}\mathrm{He}$ concentration in the system. When the $^{3}\mathrm{He}$ abundance in $^{4}\mathrm{He}$ is reduced to below $4 \times 10^{-11}$, the absorption time constant exceeds $1000$~s.

\subsubsection{Leakage area in the storage volume}\label{subsubsec:converter-leakage-area}

\begin{figure}[!htb]
\centering
\includegraphics[width=\linewidth]{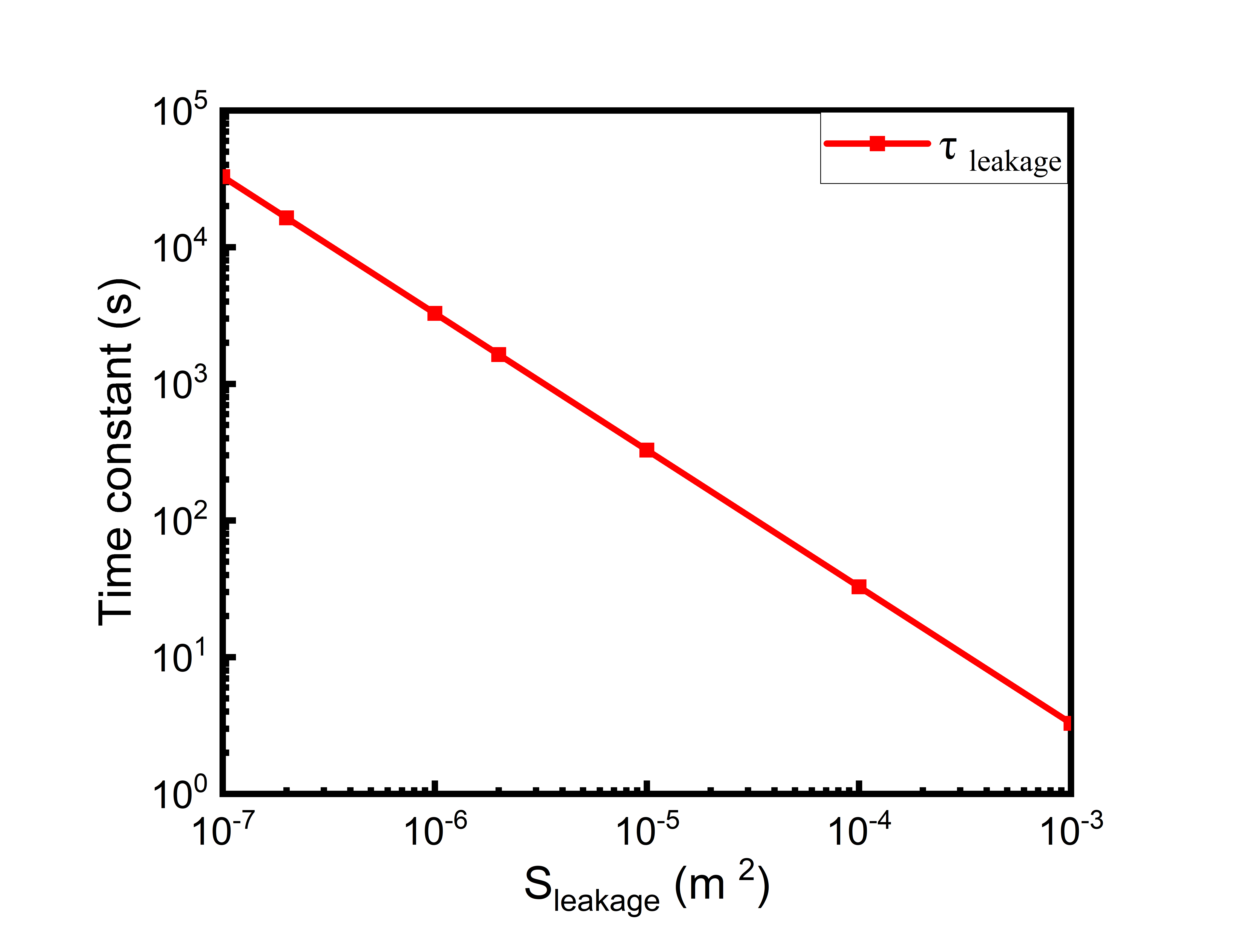}
\caption{The leakage time constant $\tau_{\mathrm{leakage}}$ as a function of the leakage area.}\label{Sleakage}
\end{figure}

At the joints of the converter walls and at the UCN shutter, gaps may exist, and defects may also occur in the coating on the converter material surface. These imperfections can lead to a finite probability of UCN loss~\cite{bib:50}, resulting in a reduction of both the storage time constant and the saturated UCN density in the converter. The leakage time constant $\tau_{\mathrm{leakage}}$ is defined as:

\begin{equation}  
\frac{1}{\tau_{\mathrm{leakage}}} =\frac{S_{\mathrm{leakage}}v}{4V} \label{tleakage}  
\end{equation}

where $S_{\mathrm{leakage}}$ represents the total area of the gaps, $v$ is the UCN velocity, and $V$ is the volume of the converter. This expression follows the same physical principle as the extraction time constant in Eq. \eqref{eq:guide-extraction-time}. Assuming a mean UCN velocity of $6~\mathrm{m}/\mathrm{s}$, the leakage time constant decreases as the gap area increases, as calculated from Eq.~\eqref{tleakage} and shown in Fig.~\ref{Sleakage}. Consequently, to maximize the saturated UCN density, the gap area must be minimized. For a typical gap area of $3~\mathrm{mm}^2$, the leakage time constant is approximately $1000~\mathrm{s}$. 
By using materials such as CYTOP \cite{bib:56} or dPS \cite{bib:71}, the leakage area in the converter can be ignored. Thus, leakage through the converter walls will be neglected in the subsequent calculations. Based on the analysis of the SUN1 experiment reported by Zimmer et al.~\cite{bib:19}, Han et al.~\cite{bib:36} estimated an effective shutter leakage area of approximately $1.73~\mathrm{mm}^2$ by comparing the simulated and measured UCN count rates with the shutter closed. Therefore, this value is adopted as the assumed shutter leakage area in the simulations of the CSNS UCN source.

\subsubsection{Converter wall materials}\label{subsubsec:converter-wall-material}

Based on previous studies and practical considerations for UCN source design, several criteria can be identified for selecting suitable wall materials \cite{bib:48}:

\begin{enumerate}
\renewcommand{\labelenumi}{(\alph{enumi})}
\item Low neutron absorption and upscattering cross sections to minimize direct losses.
\item Minimal heat load generated by the absorption of accompanying $\gamma$-rays.
\item Weak interaction with the $\gamma$-ray background to reduce radiation effects.
\item A high Fermi potential $V_0$ for internal surfaces to maximize the captured energy range of UCNs.
\item Chemical stability and non-toxicity.
\item Surfaces that are easy to clean and degas.
\item Stable surface performance at cryogenic temperatures (e.g., 0.7 K).
\item Appropriate surface roughness to limit wall losses while allowing sufficient diffuse reflection to promote UCN phase-space filling and extraction.

\end{enumerate}

The selection of the converter wall material significantly influences the wall loss time constant. According to Ref. \cite{bib:50}, the wall loss time constant $\tau_{\mathrm{wall}}$ is defined as:\begin{equation}\frac{1}{\tau_{\mathrm{wall}}} = \frac{\bar{\mu}(E_{\mathrm{UCN}}) v S}{4 V} 
\label{twall}\end{equation}where $\bar{\mu}(E_{\mathrm{UCN}})$ is the average loss probability per bounce, which is a function of the UCN energy $E_{\mathrm{UCN}}$. $S$ and $V$ denote the surface area and volume of the converter, respectively. 

\begin{figure}[!htb]
\centering
\includegraphics[width=\linewidth]{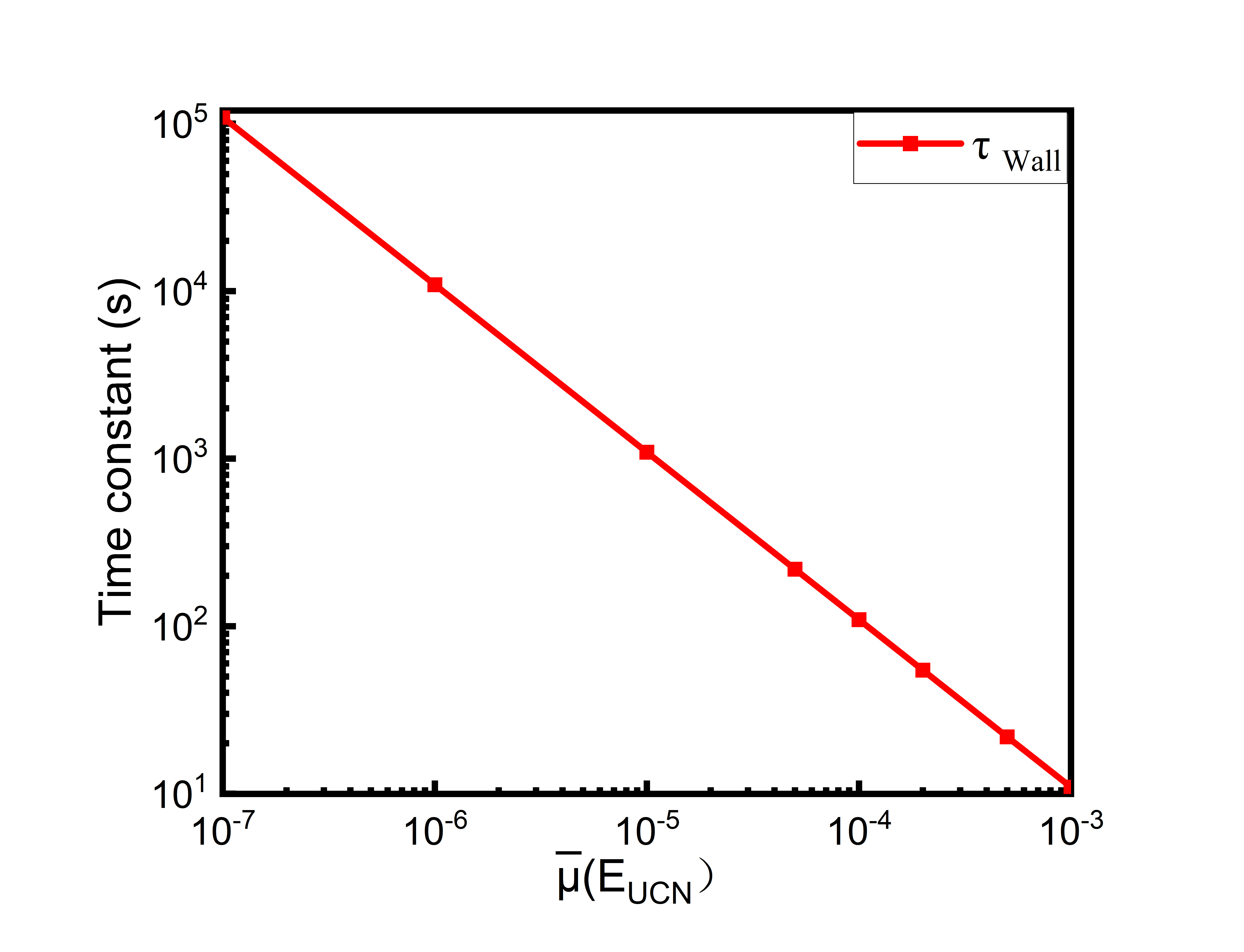}
\caption{The wall loss time constant $\tau_{\mathrm{wall}}$ as a function of the loss per bounce $\mu (E_{\mathrm{UCN}}) $.}\label{miu}
\end{figure}

As illustrated in Fig.~\ref{miu}, assuming a mean UCN velocity of $6~\mathrm{m}/\mathrm{s}$, the wall loss time constant decreases as the loss probability per bounce increases. The larger the wall loss time constant, the longer the storage duration and the higher the saturated UCN density in the converter. The wall loss time constant is a complex parameter that depends not only on the optical potential and loss rate of the wall material but also on the diffuse reflection properties of the surface. Although the optical potential and loss rate for a specific material are stable, surface impurities in real experimental environments can significantly reduce the wall loss time constant and alter the diffuse reflection probability. Additionally, wall loss is energy-dependent, with lower-energy UCNs exhibiting longer storage lifetimes than higher-energy ones \cite{bib:50}. Consequently, while a converter coated with high-optical-potential materials can store higher-energy UCNs, these neutrons remain more susceptible to loss during wall collisions. For example, experiments in Ref. \cite{bib:72} demonstrate that the wall loss time constant is approximately 600~s for UCNs with energies below 115~neV, whereas it increases to approximately 1500~s for UCNs below 50~neV.

Due to the complexity of the storage environment, it is essential to examine the impact of various wall materials on the UCN accumulation process. In this analysis, surface impurities are disregarded. The diffuse reflection probability is set to 3 \% for the simulations. A more detailed discussion regarding the impact of the converter diffuse reflection probability is provided in Section~\ref{subsubsec:extraction-guide-diffuse}.

Unlike the parametric comparison before, where an effective storage time constant of 200 s was assumed, the wall loss time constant and storage time constants listed in Table~\ref{converter-wall-material} were obtained explicitly using the iMCUCN code. The initial UCN spectrum within the superfluid $^{4}\mathrm{He}$ converter consistently follows $\mathrm{d}P_{\mathrm{source}}/\mathrm{d}E \propto \sqrt{E}$ \cite{bib:36}. The maximum initial UCN energy was adjusted according to the neutron optical potential of each converter wall material. Table \ref{converter-wall-material} summarizes the wall loss time constants, total storage time constants, UCN production rates per unit volume, and saturated UCN densities for various candidate converter wall materials. 

The wall loss time constants listed in Column 2 of Table~\ref{converter-wall-material} were calculated using iMCUCN for a closed rectangular converter with dimensions of $100 \times 7 \times 7~\mathrm{cm}^3$, without the extraction system. In each simulation, 10,000 UCNs were initially generated inside the converter, and all loss mechanisms in Eq.~\eqref{ttotal} except wall losses were disabled, so that only wall losses associated with the different converter materials were considered. The wall loss time constant $\tau_{\mathrm{wall}}$ was determined by fitting the number of UCNs remaining in the converter as a function of time using $N(t)=N_0\exp(-t/\tau_{\mathrm{wall}})$.

The total storage time constants listed in Column 3 of Table~\ref{converter-wall-material} were then calculated using iMCUCN for the full converter geometry, including the UCN extraction system. All UCN loss mechanisms discussed in Section~\ref{subsec:converter} were included in these simulations.
The extraction system consists of an initial vertical guide extending 10 cm upward from the converter, with a shutter positioned midway along this section. The guide then includes a 7 cm horizontal section, followed by a second vertical section of 15 cm leading to the detector. The extraction guides are made of stainless steel and have an inner diameter of 23 mm.
The superfluid $^{4}\mathrm{He}$ temperature was set to $0.7~\mathrm{K}$, corresponding to an upscattering time constant of $\tau_{\mathrm{up}}=1247~\mathrm{s}$. The $^{3}\mathrm{He}/^{4}\mathrm{He}$ abundance ratio was set to $4\times10^{-11}$, corresponding to an absorption time constant of $\tau_{\mathrm{abs}}\approx1000~\mathrm{s}$. An effective leakage area of $1.73~\mathrm{mm^2}$ was adopted. The cold-neutron shutter was opened at $t=18~\mathrm{s}$ and closed at $t=804~\mathrm{s}$, corresponding to a UCN production duration of $786~\mathrm{s}$. A total of $2.5\times10^{6}$ UCNs were generated inside the converter during this period, and the total storage time constant $\tau$ was determined by fitting the UCN accumulation curve using Eq.~\eqref{rho-t}. The shutter remained closed throughout the storage simulations. Thus, the extraction-guide and shutter geometries were included in the storage configuration, but no intentional UCN extraction was performed.

The UCN production rates per unit volume in Column 4  of Table~\ref{converter-wall-material} were calculated based on Eq.~\eqref{P}. For the Be-coated converter, the production rate of $P=2.46~\mathrm{UCN/cm^3/s}$ was obtained from the selected $100\times7\times7~\mathrm{cm^3}$ converter configuration, as shown in Fig.~\ref{converter_length}. The influence of the converter wall coatings, with thicknesses on the order of hundreds of nanometers, on cold-neutron transport was assumed to be negligible. The production rates for the other wall materials were then scaled according to the trapping energy, $E_{\mathrm{trap}} = V_{\mathrm{wall}} - V_{^{4}\mathrm{He}}$, using the relation $P \propto E_{\mathrm{trap}}^{3/2}$. Finally, the saturated UCN densities in Column 5  of Table~\ref{converter-wall-material} were obtained from Eq.~\eqref{rho}, using the production rates in Column 4 and the total storage time constants in Column 3. For example, for Be, $P = 2.46~\mathrm{UCN/cm^3/s}$ and $\tau = 130~\mathrm{s}$ give $\rho_{\mathrm{sat}} \approx 320~\mathrm{UCN/cm^3}$.

As shown in Table~\ref{converter-wall-material}, Be provides the highest predicted saturated UCN density among the wall materials considered. DLC is an emerging material that is currently considered unsuitable for the initial phase of the CSNS UCN source due to practical considerations. Other potential candidates include Fomblin and CYTOP. Previous sources, such as SUN1 and SUN2, utilized BeO and Be as converter wall materials, respectively; however, they exhibited storage time constants below 100~s, likely due to surface impurities or structural gaps \cite{bib:18,bib:19,bib:20}. While coating these Be and BeO vessels with Fomblin grease significantly enhanced the time constant, Fomblin coatings are prone to flow over time, leading to non-uniform thickness and regional clumping. Consequently, CYTOP has been investigated for the construction of SuperSUN \cite{bib:21,bib:56}. Considering their availability, cost, and practical applicability, Be and CYTOP are considered the primary candidate wall materials for the initial CSNS UCN source design.

\begin{table*}[!htb]
\centering
\caption{Physical parameters of candidate wall materials and their corresponding simulated wall loss time constants under identical conditions.}

\setlength{\tabcolsep}{10pt}
\renewcommand{\arraystretch}{1.0}

\begin{tabular}{ccccc}
\toprule
Material
&
\begin{tabular}{c}
Wall loss time\\
constant (s)
\end{tabular}
&
\begin{tabular}{c}
Storage time\\
constant (s)
\end{tabular}
&
\begin{tabular}{c}
Production rate per unit volume\\
(UCN/cm$^{3}$/s)
\end{tabular}
&
\begin{tabular}{c}
UCN density\\
(UCN/cm$^{3}$)
\end{tabular}
\\
\midrule
$^{58}\mathrm{Ni}$ & 93 & 72 & 3.88 & 279 \\
DLC & 481 & 144 & 1.97 & 284 \\
BeO & 307 & 115 & 2.60 & 299 \\
Be & 382 & 130 & 2.46 & 320 \\
SS & 131 & 78 & 1.45 & 113 \\
Fomblin & 475 & 140 & 0.65 & 91 \\
CYTOP & 419 & 134 & 0.65 & 87 \\
dPS & 406 & 135 & 1.17 & 158 \\
\bottomrule
\end{tabular}
\label{converter-wall-material}
\end{table*}

\subsection{The extraction}\label{subsec:extraction}

The extraction system is a critical factor determining the number of UCNs available for experiments. This section investigates the impact of extraction geometry, the extraction-guide wall material, the guide radius, and the diffuse reflection probability on the overall UCN extraction efficiency. This efficiency is defined by Eq. \eqref{Trasmission}, which characterizes the transport performance from the converter to the experiments.

\subsubsection{The extraction method}\label{subsubsec:extraction-method}

The extraction method is a crucial factor in enhancing the availability of UCNs for experimental use. In this section, four extraction geometries and their respective extraction efficiencies are explored, drawing inspiration from concepts presented in Ref. \cite{bib:73}. 
The simulations were based on the Be-coated converter configuration used to calculate the total storage time constants in Column 3 of Table~\ref{converter-wall-material}, with only the extraction geometry varied. Unlike the storage simulations, the initial UCN population was generated at $t=0$, and the shutter remained open throughout the extraction process, without a preceding accumulation stage. Therefore, UCN leakage associated with the closed shutter was not included in the extraction simulations.

\begin{figure}[!htb]
  \centering
  \subfloat[Down extraction: $T = 38.0\%$]{\includegraphics[width=0.45\linewidth]{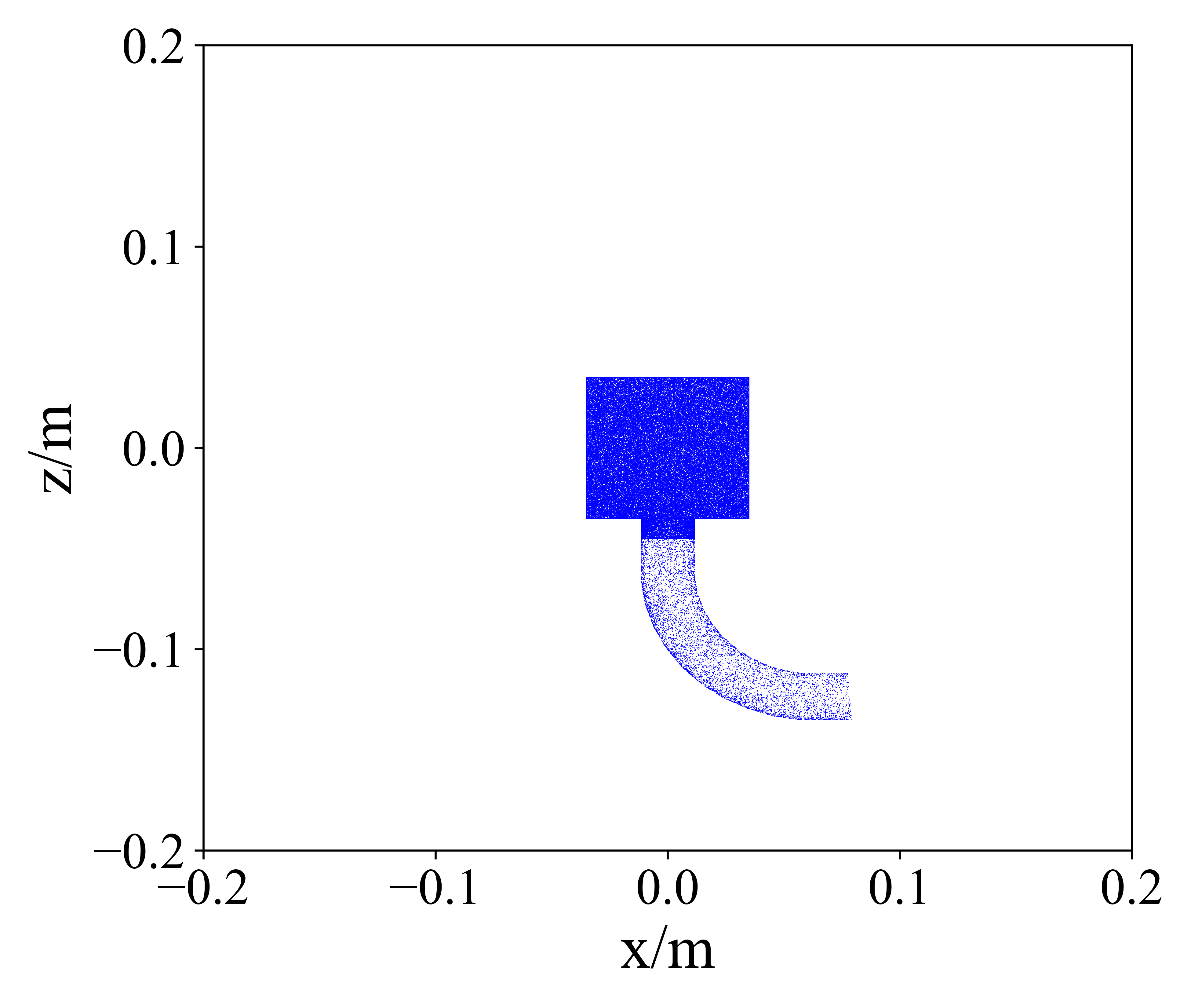}\label{extract3}}
  \hfill
  \subfloat[Side extraction: $T = 40.0\%$]{\includegraphics[width=0.45\linewidth]{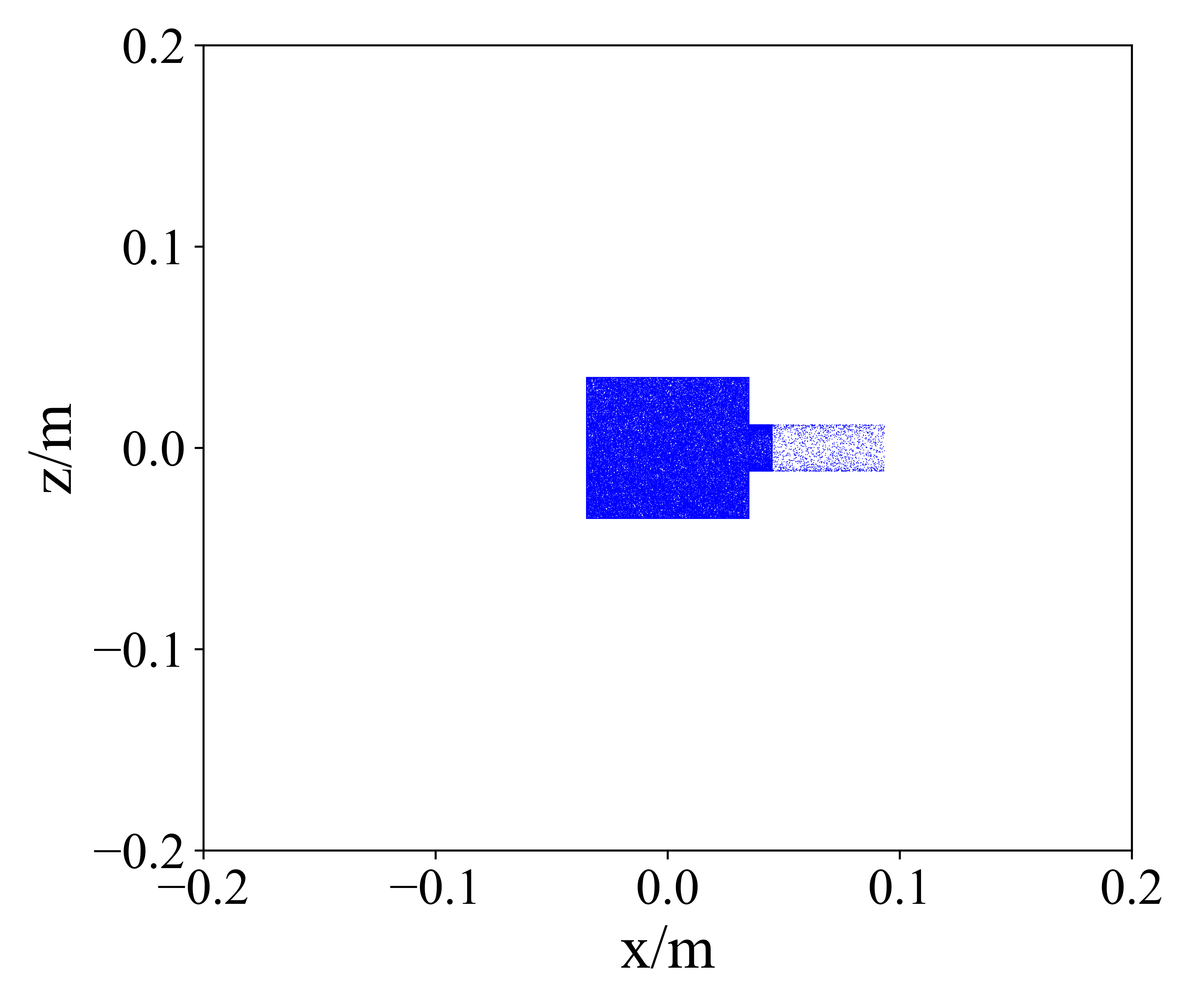}\label{extract5}}
  \par\medskip
  \subfloat[Up-bend-down extraction: $T = 42.0\%$]{\includegraphics[width=0.45\linewidth]{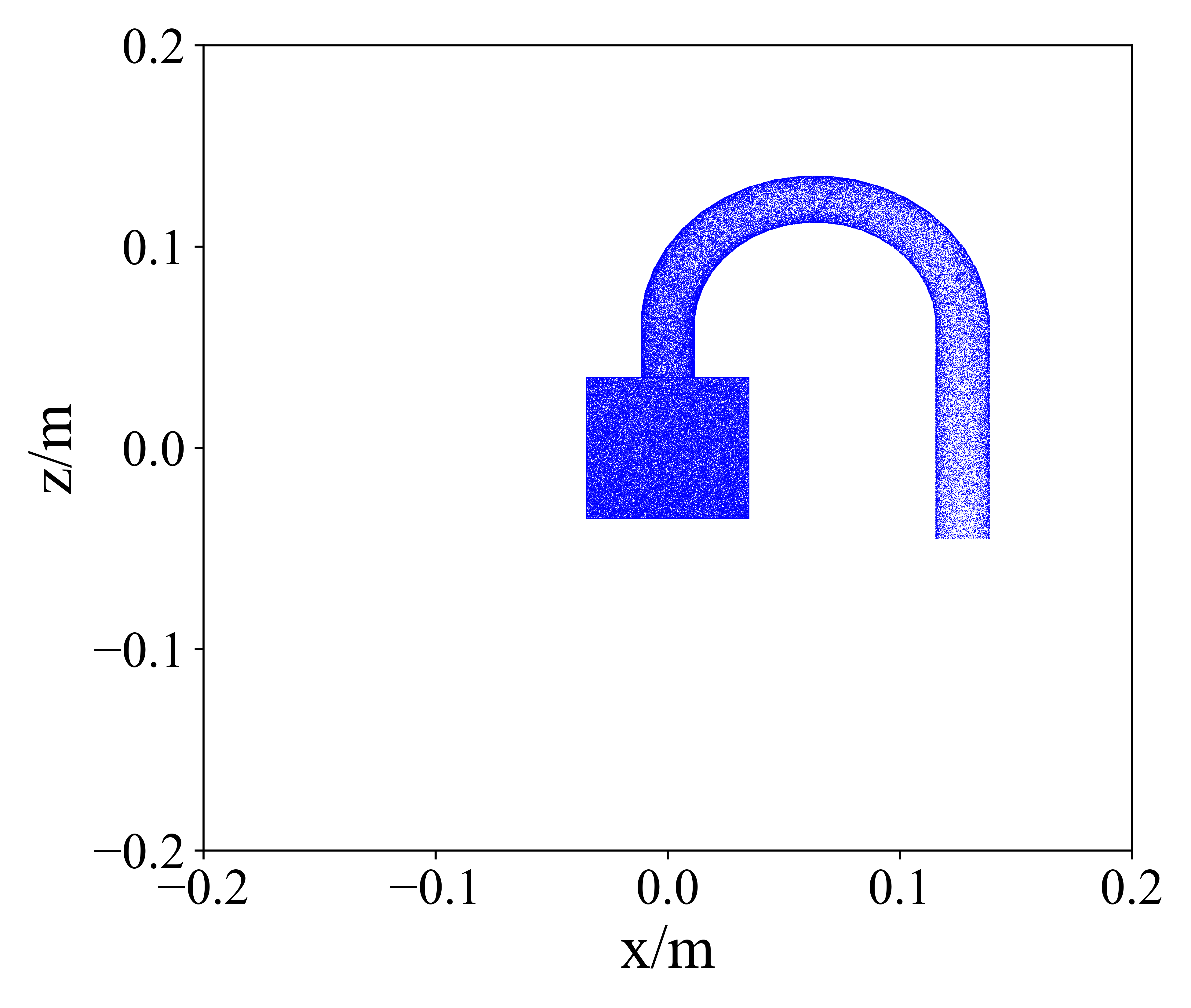}\label{extract2}}
  \hfill
  \subfloat[Up extraction: $T = 82.0\%$]{\includegraphics[width=0.45\linewidth]{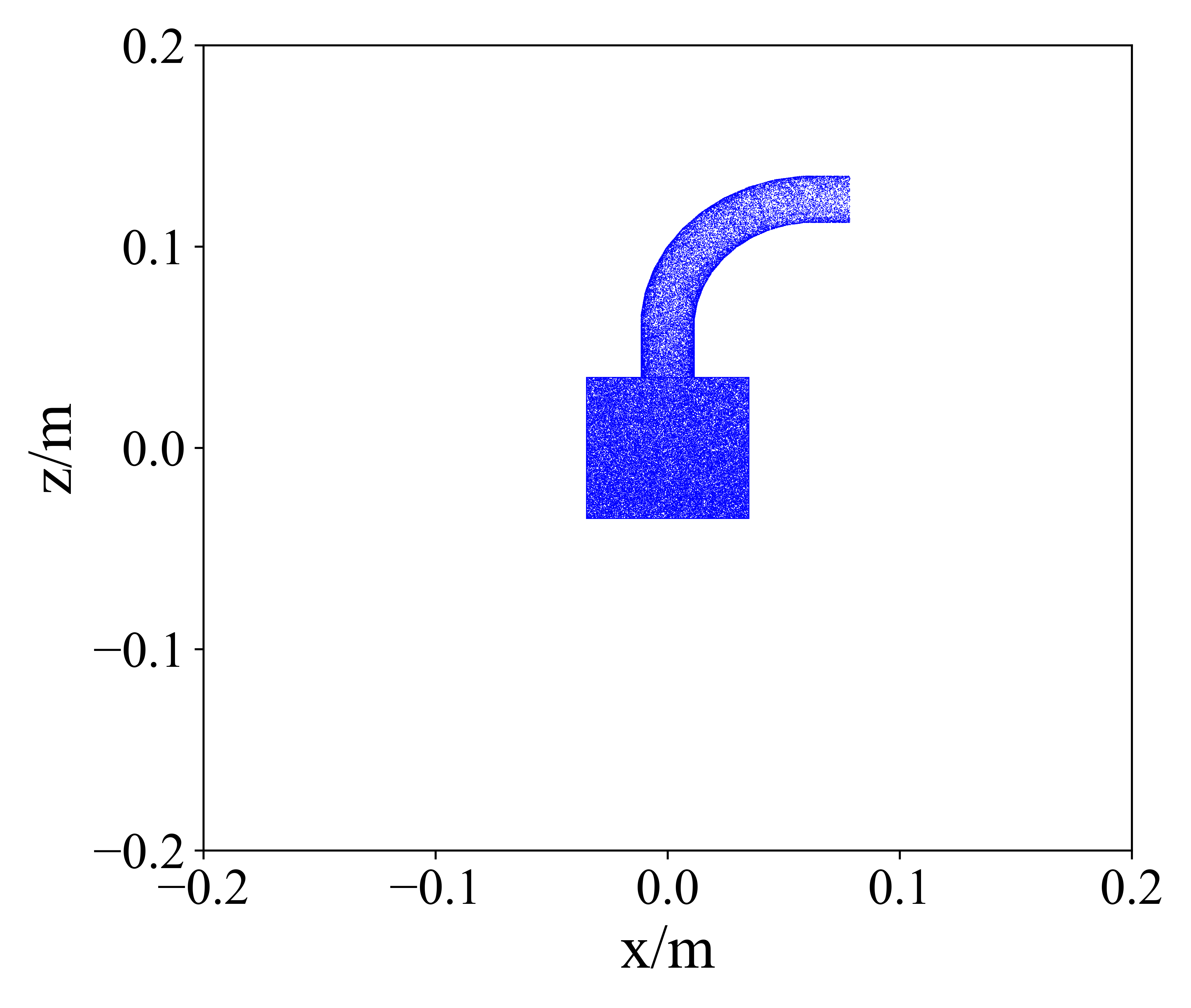}\label{extract1}}
  
  \caption{The four extraction methods and their extraction efficiencies, including down extraction, side extraction, up-bend-down extraction, and up extraction.}
  \label{extraction}
\end{figure}

As illustrated in Fig.~\ref{extraction}, the up-extraction method achieves the highest extraction efficiency of 82.0\%, significantly exceeding those of the other three configurations (38.0\%--42.0\%), primarily because it does not require a separation foil. In contrast, the down-, side-, and up-bend-down extraction methods require a separation foil to confine the superfluid $^{4}\mathrm{He}$, which introduces additional UCN losses due to absorption and upscattering and thereby reduces the extraction efficiency.

\begin{figure}[htbp]
  \centering
  \subfloat[$T = 80.4\%$]{\includegraphics[width=0.45\linewidth]{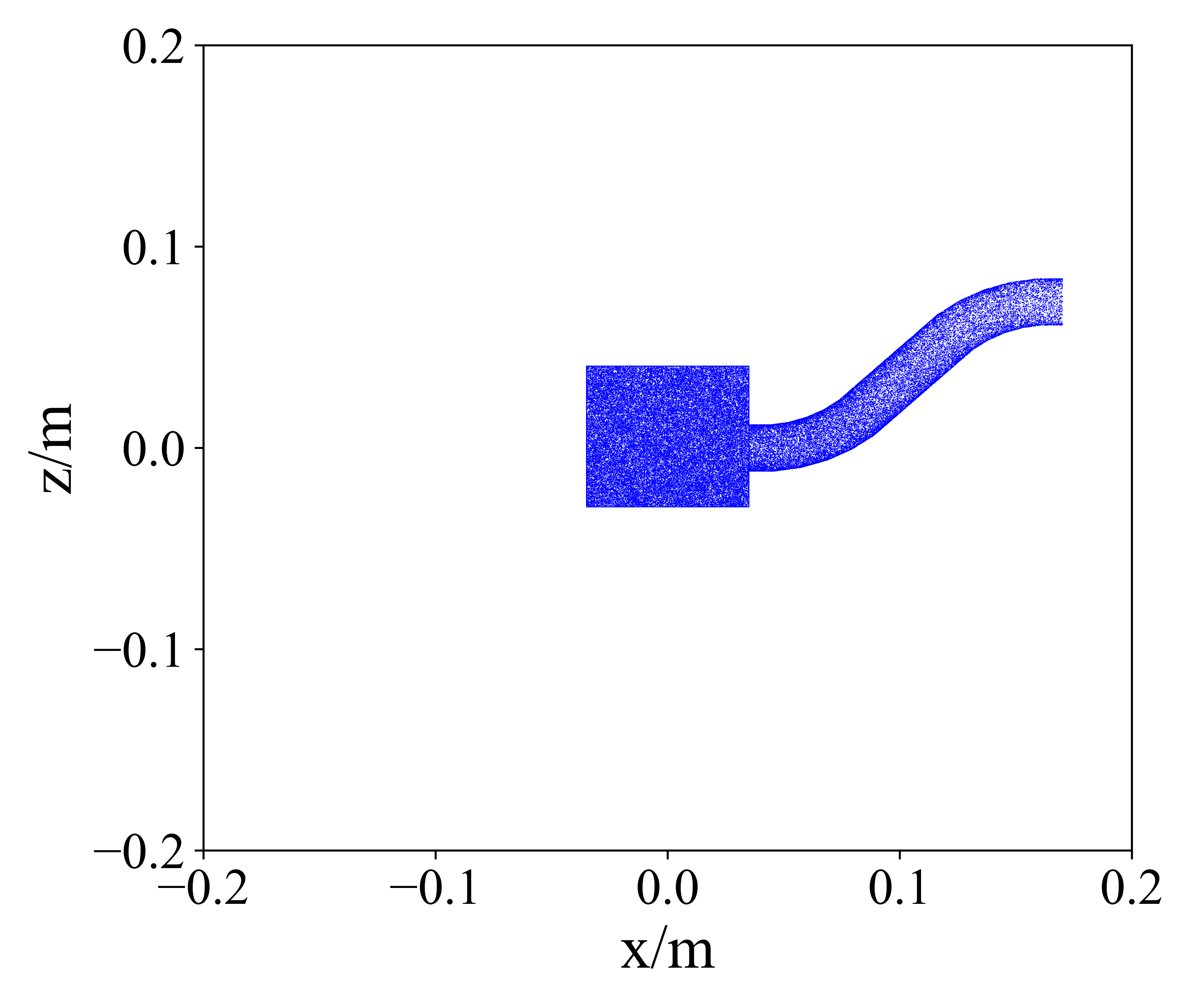}\label{extract6}}
  \hfill
  \subfloat[$T = 81.0\%$]{\includegraphics[width=0.45\linewidth]{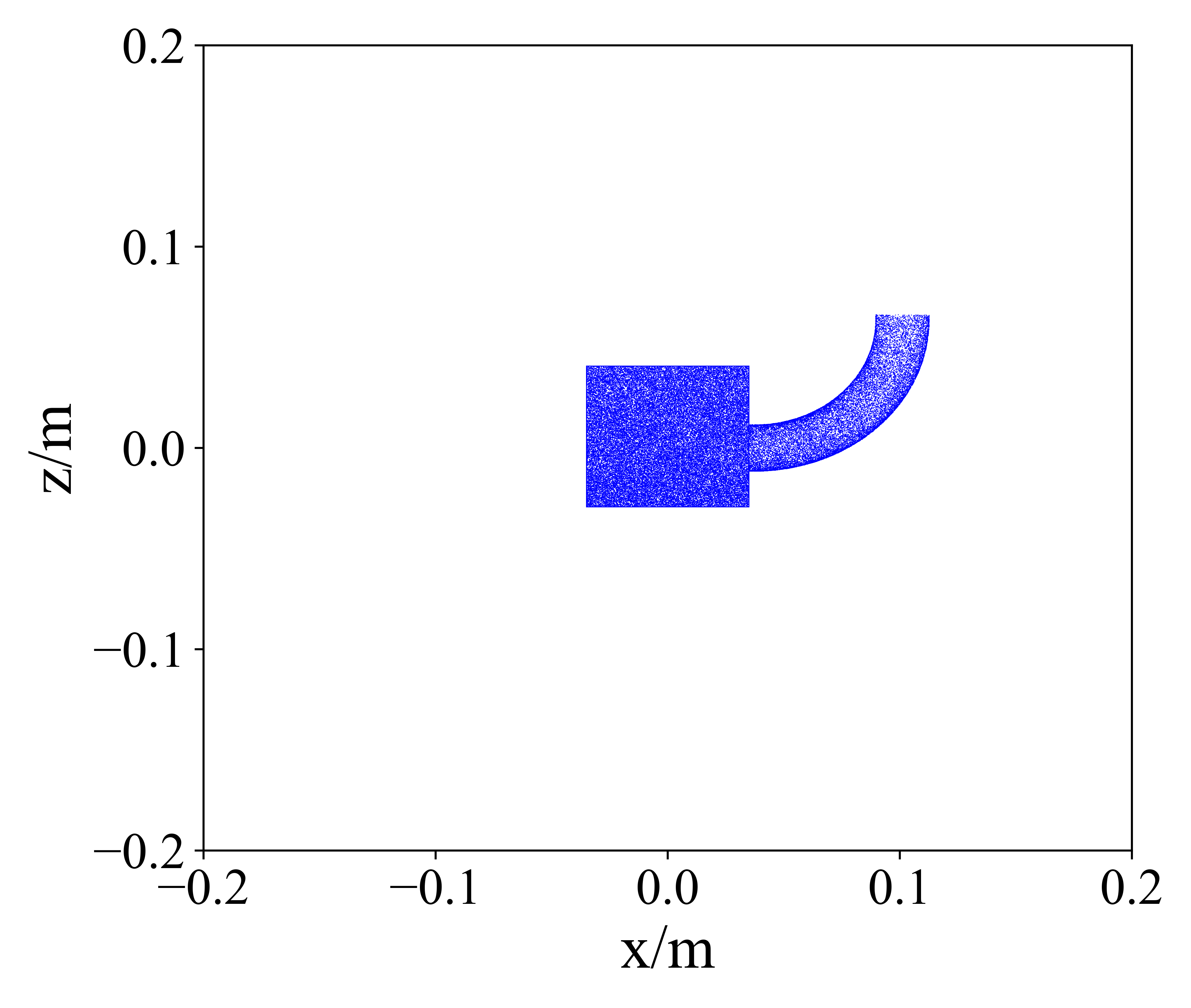}\label{extract7}}
  \par\medskip
  \subfloat[$T = 75.7\%$]{\includegraphics[width=0.45\linewidth]{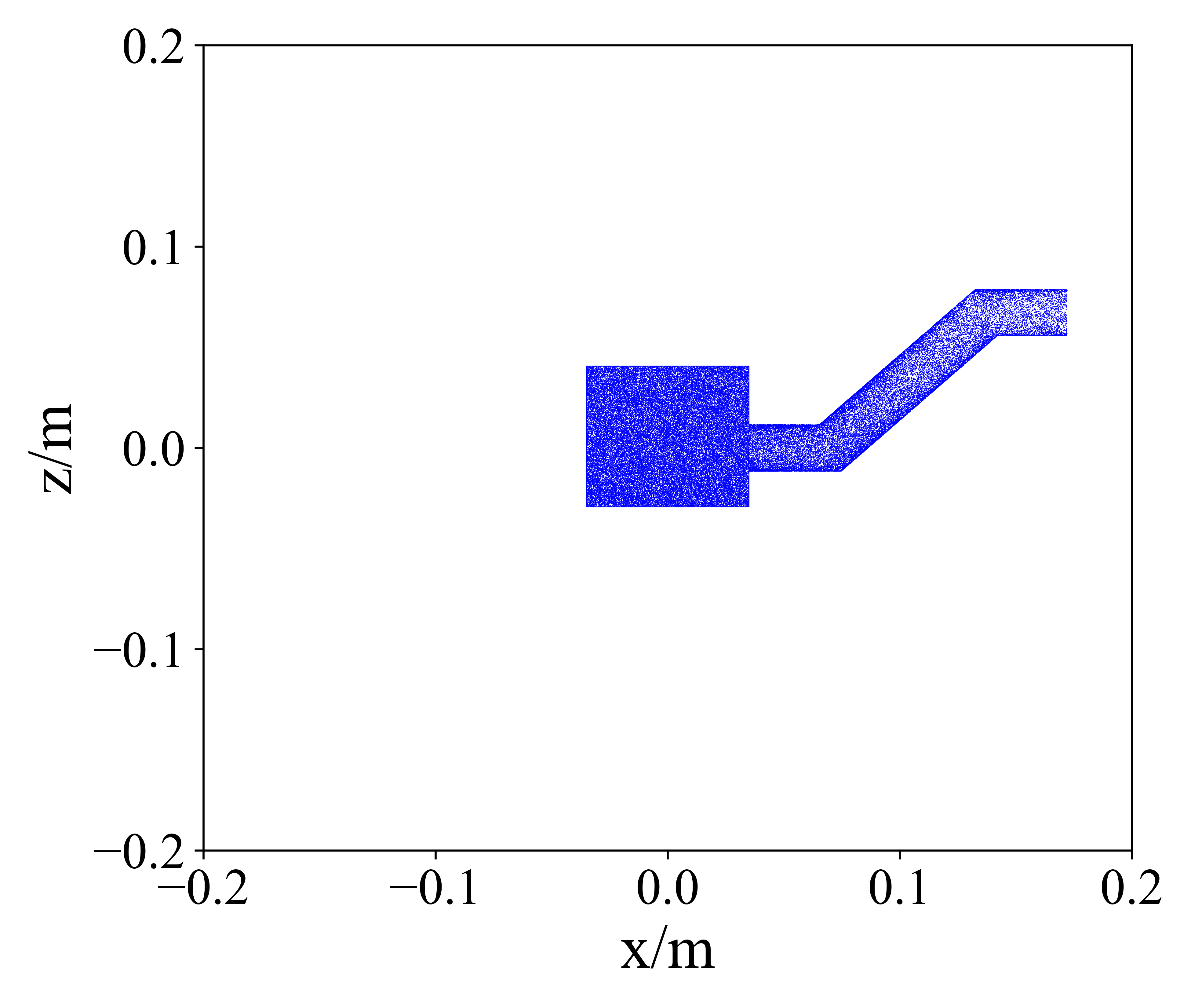}\label{extract8}}
  \hfill
  \subfloat[$T = 80.0\%$]{\includegraphics[width=0.45\linewidth]{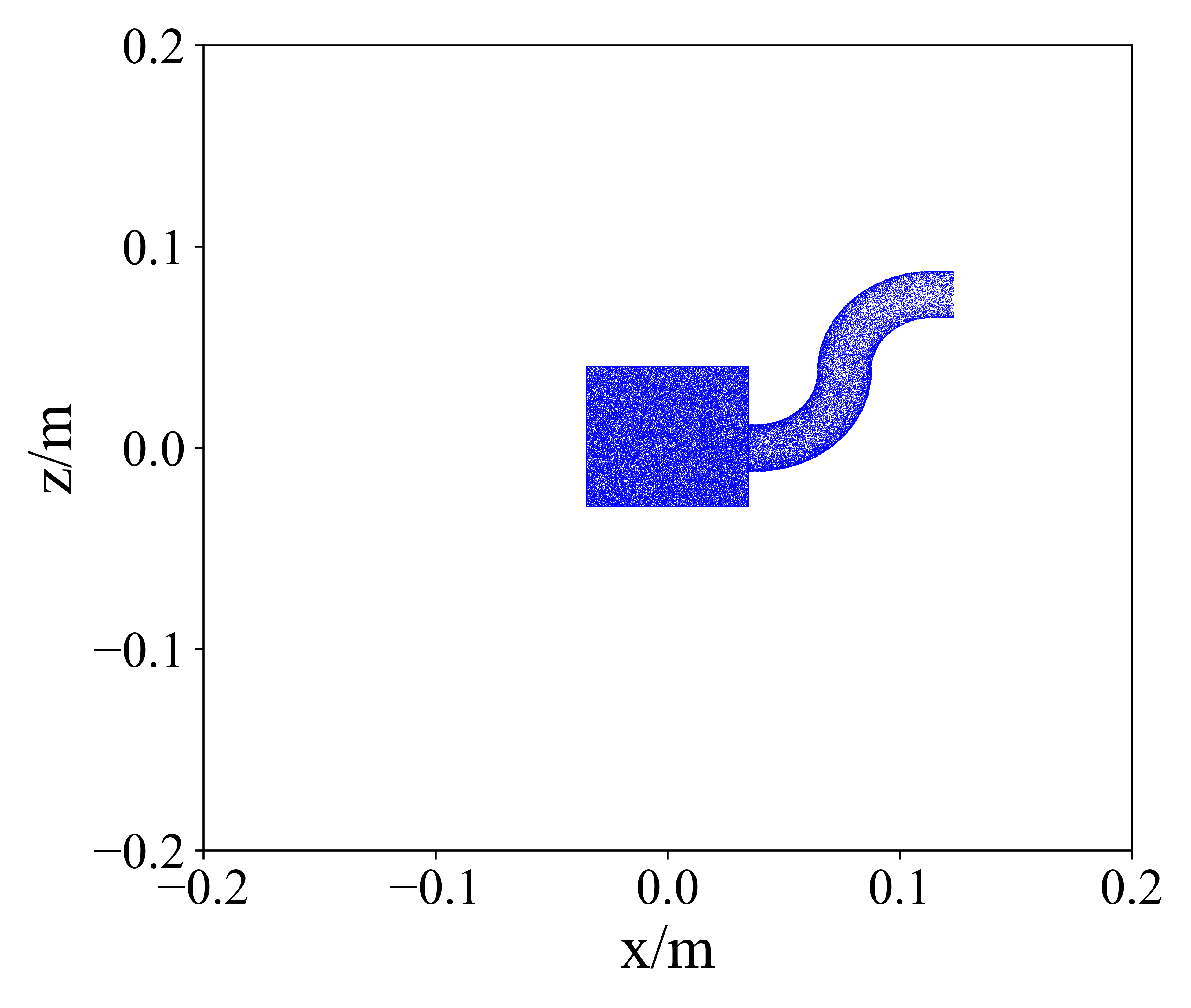}\label{extract9}}
  \par\medskip
  \subfloat[$T = 77.1\%$]{\includegraphics[width=0.45\linewidth]{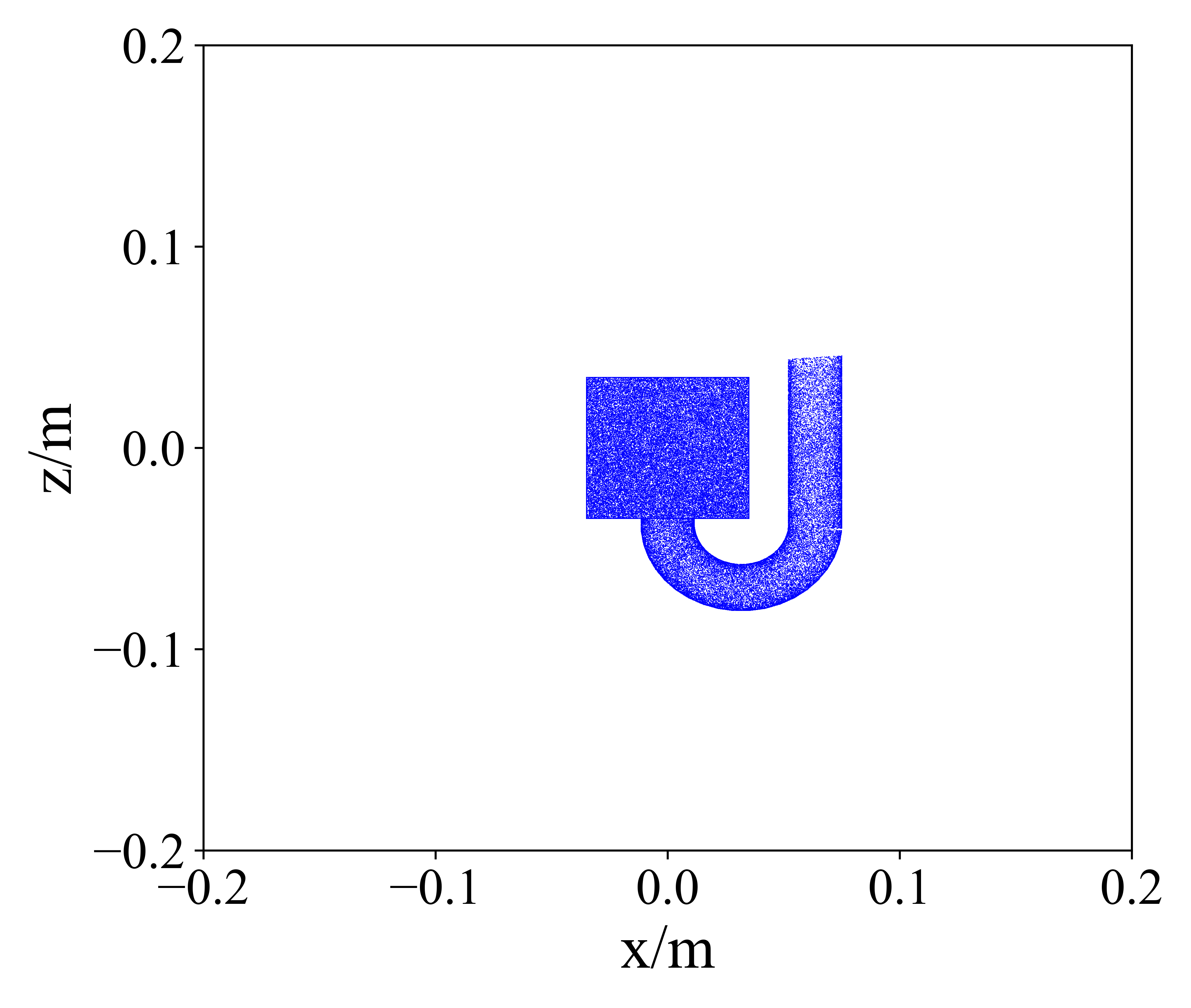}\label{extract10}}
  
  \caption{The different types of extraction geometries and their extraction efficiencies.}
  \label{up-extraction}
\end{figure}

As illustrated in Fig.~\ref{up-extraction}, several extraction-guide geometries were investigated to enable foil-free UCN extraction. In these configurations, the guide geometry was designed to raise the extraction path above the superfluid $^{4}\mathrm{He}$ level in the converter, thereby preventing bulk liquid helium from overflowing. These configurations represent idealized preliminary models intended to investigate the influence of guide geometry on UCN extraction efficiency rather than fully developed engineering designs. The calculated extraction efficiencies range from 75.7\% to 81.0\%, showing relatively small variations among the configurations. Owing to the relatively short guide lengths, UCNs are expected to undergo fewer wall collisions, thereby limiting additional wall losses during transport.

Consequently, the extraction efficiency for these relatively short extraction guides is strongly influenced by gravity, while the differences among the guide geometries remain relatively small. Based on these results, the UCN source at CSNS will utilize the up-extraction method. However, in practical operation, the challenges associated with windowless up-extraction include: (1) a high heat load to the superfluid $^{4}\mathrm{He}$ due to thermal radiation through the guides, and (2) the continuous freezing of contaminants on the cold extraction guide surfaces, which may reduce the extraction efficiency over time \cite{bib:61}.

\subsubsection{Extraction guide radius}\label{subsubsec:extraction-guide-radius}

\begin{figure}[!htb]
\centering
\includegraphics[width=0.45\textwidth]{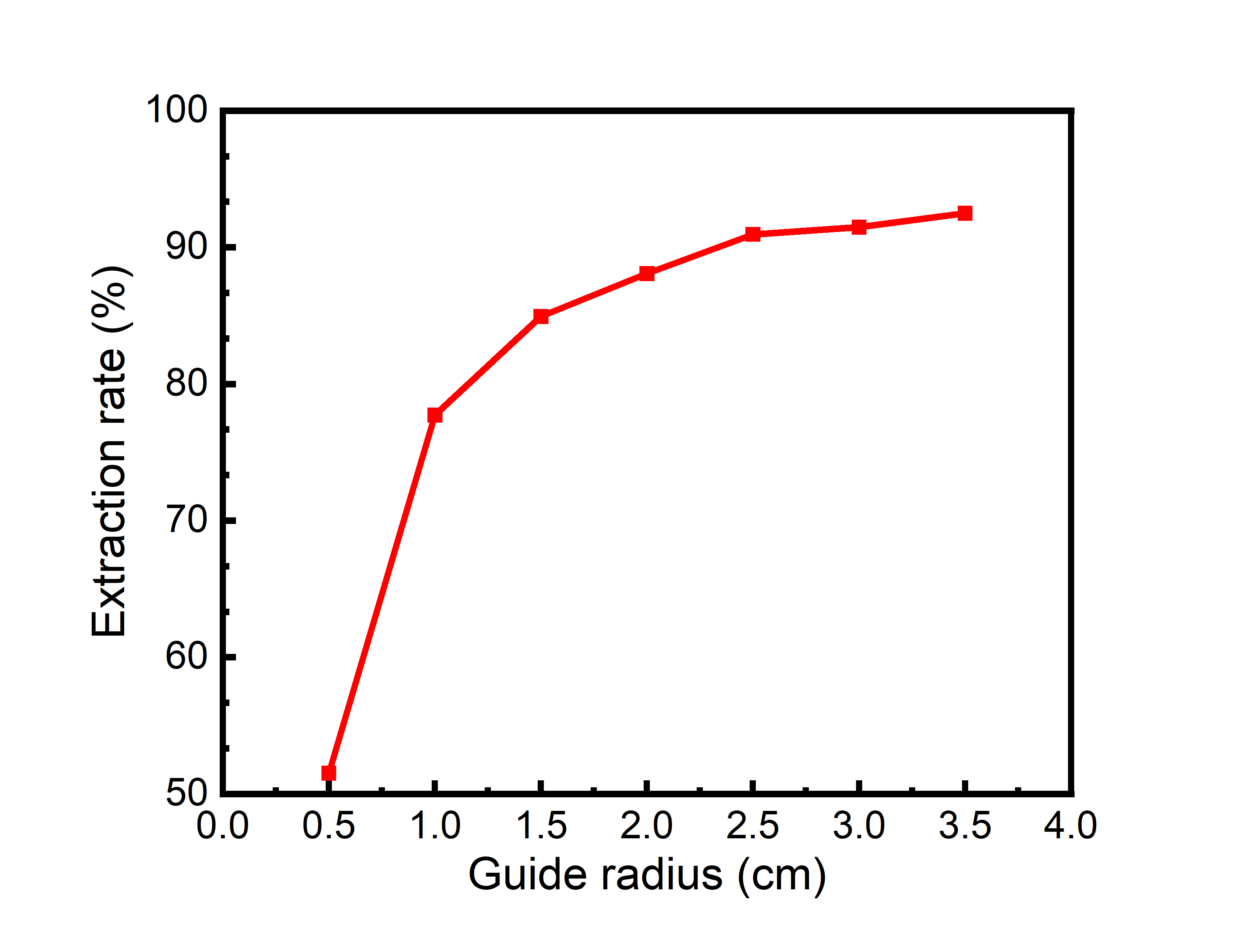}
\caption{The extraction efficiency as a function of the extraction guide radius.}
\label{guide-radius}
\end{figure}

The effect of the guide radius on the extraction efficiency for the up-extraction configuration shown in Fig.~\ref{extract1} was examined. As illustrated in Fig.~\ref{guide-radius}, the extraction efficiency increases from approximately 51.5\% to 92.5\% as the guide radius increases from 0.5 to 3.5~cm, while the other parameters are held constant. The UCN extraction process can be conceptualized as neutrons escaping through an aperture defined by the guide's cross section. While a larger guide area $A$ enhances the UCN extraction efficiency, it concurrently increases the heat load $Q$ on the superfluid $^{4}\mathrm{He}$ converter. Consequently, a full engineering optimization of the guide radius must consider the $^{3}\mathrm{He}$ refrigerator's cooling power to maintain the necessary converter temperature $T_{\mathrm{c}}$. Based on the simulated extraction efficiency, a guide radius of 2.5~cm was provisionally selected for the preliminary design.

\subsubsection{The guide material}\label{subsubsec:guide-material}
\begin{figure}[!htb]
\centering
\includegraphics[width=0.45\textwidth]{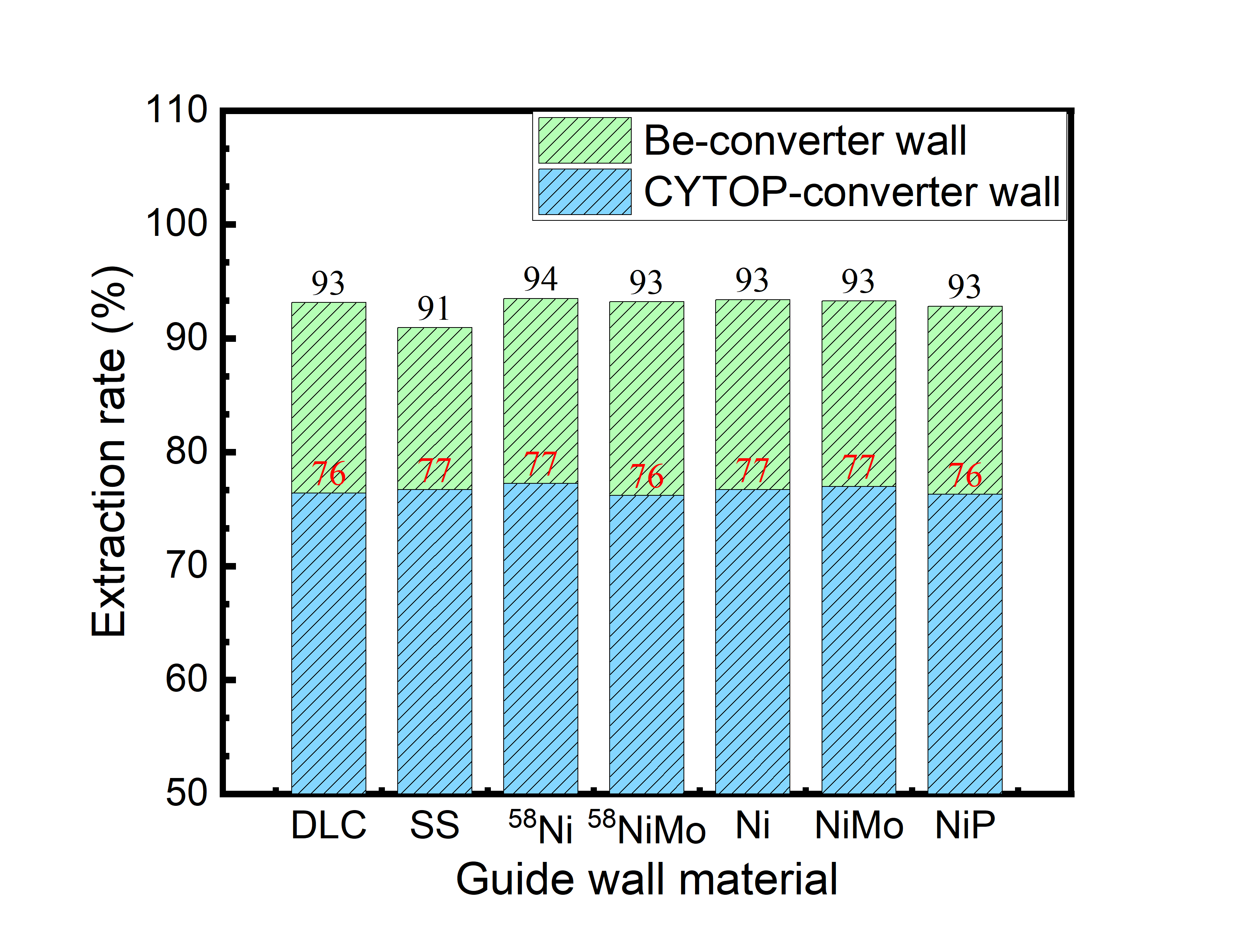}
\caption{The UCN extraction efficiencies of different guide materials when the converter wall material is CYTOP or Be.}
\label{guide-material}
\end{figure}

Different UCN sources utilize various extraction guide materials. The extraction efficiencies for systems with converters coated in Be and CYTOP were evaluated using the selected up-extraction configuration with a guide radius of 2.5~cm, as shown in Fig.~\ref{guide-material}. During the 20~cm vertical up-extraction process, approximately 20~neV of UCN kinetic energy is transformed into gravitational potential energy.

The extraction efficiency for the CYTOP-coated converter ($V_0 \approx 115$~neV) is significantly lower than that for the Be-coated converter ($V_0 \approx 252$~neV), because the lower trapping potential of CYTOP restricts the stored UCNs to a lower energy range, resulting in a larger fraction being unable to overcome the gravitational potential during upward extraction. While the extraction efficiency for a CYTOP-coated converter could be further improved by decreasing the vertical length of the extraction guide, such a modification must be carefully balanced against practical engineering constraints.

For the CYTOP-coated converter, the extraction efficiencies are nearly independent of the guide material, as the optical potentials of all considered guides exceed the energies of the stored UCNs. In contrast, for the Be-coated converter, the extraction efficiency decreases slightly when stainless steel is used due to its lower optical potential ($V_0 \approx 183$~neV), which causes additional losses of higher-energy UCNs during transport.

\subsubsection{Diffuse reflection probability of the extraction guide}\label{subsubsec:extraction-guide-diffuse}

UCNs interact with material surfaces through a combination of specular and diffuse reflections. During the extraction process, a high diffuse reflection probability ($f$) significantly extends the residence time of UCNs within the guide by increasing their effective path length, thereby amplifying the probability of absorption and upscattering losses. The effects of diffuse reflection were investigated using the selected up-extraction configuration with a guide radius of 2.5~cm and stainless steel (SS) as the guide material.

\begin{figure}[!htb]
\centering
\includegraphics[width=0.45\textwidth]{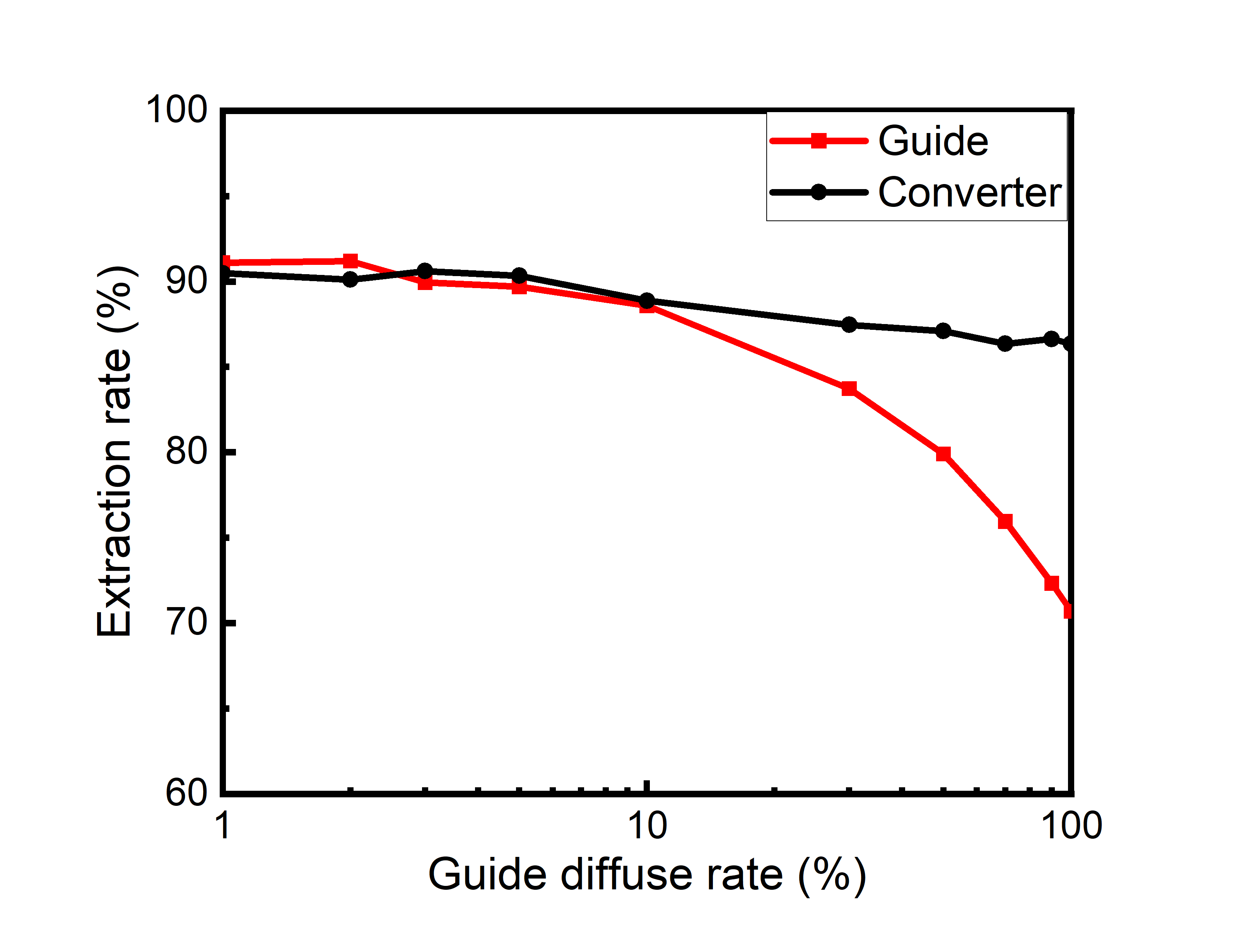}
\caption{The extraction efficiency as a function of the diffuse reflection probabilities of the guide and converter.}
\label{guide-diffuse}
\end{figure}

As illustrated in Fig. \ref{guide-diffuse}, the extraction efficiency exhibits a pronounced downward trend when the diffuse reflection probability exceeds 10\%. This behavior can be analytically understood using diffusion theory, where the presence of a specular reflection component ($1-f$) enhances the diffusion coefficient ($D$) by a factor of $(2-f)/f$ \cite{bib:50}. For typical electropolished guides with a diffuse reflection probability $f$ of 3\% \cite {bib:36}, the diffusion coefficient is amplified by a factor of approximately 65.7 compared to the purely diffuse case ($f=1$), drastically increasing the axial conductance of the guide. The sharp decrease in efficiency beyond the 10\% threshold reflects the exponential nature of UCN survival, which follows a decay law of $(1-\mu)^n$, where $\mu$ denotes the loss probability per bounce. At higher diffuse reflection probabilities, the rapid accumulation of collision events ($n$) significantly amplifies the cumulative wall losses. In the simulated configuration, the decrease above a diffuse reflection probability of approximately 10\% indicates that increased wall interactions reduce the extraction efficiency. This underscores the critical necessity of maintaining superior surface smoothness in long-distance UCN guides.

In contrast, the impact of the converter wall's diffuse reflection probability on the extraction efficiency is relatively marginal. Increasing the converter's diffuse reflection probability from 0\% to 100\% results in a reduction of approximately 5\% in extraction efficiency. 
This is because moderate diffuse reflection can randomize the trajectories of UCNs trapped in long-lived quasi-stable orbits, thereby increasing their probability of escaping the converter volume and entering the extraction guide~\cite{bib:73}. This observation aligns with previously reported findings in the literature~\cite{bib:26}.

\section{The physical design of the superfluid $^{4}\mathrm{He}$ UCN source at CSNS}\label{sec:final-source-design}

Based on the analysis of the parameters in Section~\ref{sec:design-factors}, a preliminary physical design for the CSNS UCN source is proposed. The design features a rectangular converter with dimensions of \SI{100}{\cm} (length) $\times$ \SI{7}{\cm} (width) $\times$ \SI{7}{\cm} (height), filled with superfluid $^{4}\mathrm{He}$. The inner surfaces of the converter are coated with either CYTOP or Be. UCNs are extracted upward through an extraction guide with a radius of \SI{2.5}{\cm}, utilizing either $^{58}\mathrm{Ni}$ or polished stainless steel for the inner coating. In this initial stage, a moderator-reflector system is excluded, and the diffuse reflection probability for both the guide and the converter is set to 3 \%. The detailed design parameters are summarized in Table~\ref{tab:design_params}.

\begin{table*}[htbp]
\centering
\caption{Physical design parameters of the superfluid $^{4}\mathrm{He}$ UCN source at CSNS.}
\label{tab:design_params}

\begin{tabular}{ll}
\toprule
\textbf{Configuration} & \textbf{Parameter} \\
\midrule

Cold neutron integrated intensity
& $6.71 \times 10^{10}\,\mathrm{n/s}$ (0.10--12.00 \AA) \\

Differential cold neutron flux at \SI{8.9}{\angstrom}
& $3.58 \times 10^{7}$ \text{ n cm$^{-2}$ s$^{-1}$ \AA$^{-1}$} \\

Converter material
& Superfluid $^{4}\mathrm{He}$ \\

Converter dimensions
& \SI{100}{\cm} (length) $\times$ \SI{7}{\cm} (width) $\times$ \SI{7}{\cm} (height) \\

Converter temperature
& $0.7\,\mathrm{K}$ (corresponding to $\tau_{\mathrm{up}} = 1247\,\mathrm{s}$) \\

$^{3}\mathrm{He}/^{4}\mathrm{He}$ abundance ratio
& $4 \times 10^{-11}$ (corresponding to $\tau_{\mathrm{abs}} \approx 1000\,\mathrm{s}$) \\

Converter wall material
& Be / CYTOP \\

Diffuse reflection probability of converter
& $3\%$ \\

Diffuse reflection probability of guide
& $3\%$ \\

Extraction method
& Up extraction \\

Extraction guide radius
& $2.5\,\mathrm{cm}$ \\

Extraction guide material
& Polished stainless steel / $^{58}\mathrm{Ni}$ \\

\bottomrule
\end{tabular}
\end{table*}

Based on the incident cold-neutron energy spectrum shown in Fig.~\ref{fig:IncidentCN} and the physical design parameters summarized in Table~\ref{tab:design_params}, the UCN production rates are 12,055 and 3,185~UCN/s for the Be- and CYTOP-coated converters, respectively, taking into account their different trapping potentials.

Using the design parameters in Table~\ref{tab:design_params}, UCN production and storage simulations were performed for the Be- and CYTOP-coated converters. To compare their storage performance under identical UCN input conditions, both simulations employed an input UCN production rate of 12,055~UCN/s and an initial energy distribution of $\mathrm{d}P_{\mathrm{source}}/\mathrm{d}E \propto \sqrt{E}$, with a maximum initial UCN energy of 233.5~neV. The incident cold-neutron beam was switched on at $t=18~\mathrm{s}$ and switched off at $t=804~\mathrm{s}$, corresponding to a production duration of 786~s and a total of $9.48\times10^6$ generated UCNs for each converter. The simulations accounted for UCN losses due to wall interactions, upscattering in superfluid $^{4}\mathrm{He}$, neutron absorption by residual $^{3}\mathrm{He}$, neutron $\beta$ decay, and leakage through the closed shutter, with an effective leakage area of $1.73~\mathrm{mm}^2$. 

\begin{figure}[!htb]
\centering
\includegraphics[width=0.45\textwidth]{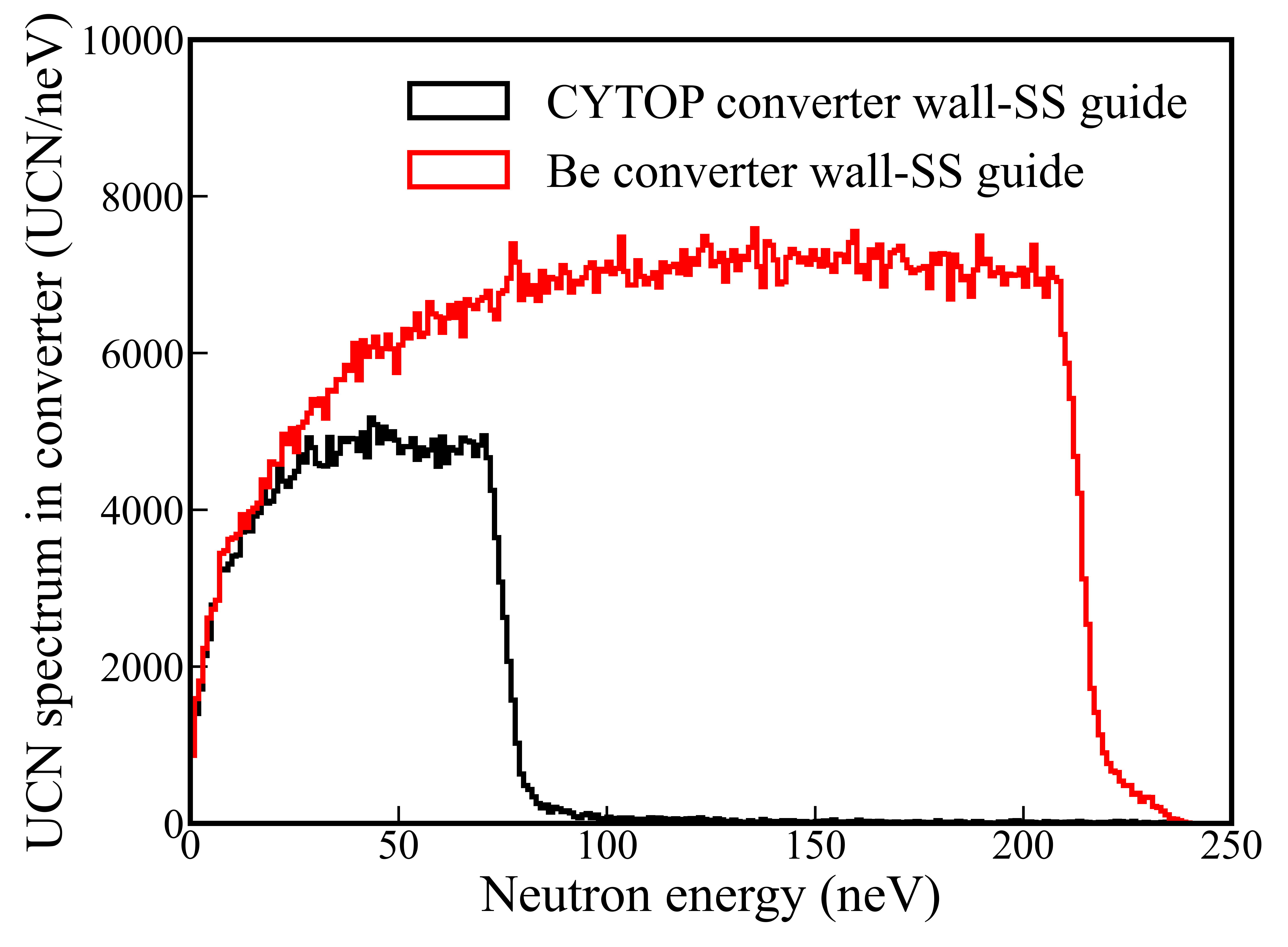}
\caption{UCN energy spectra in CYTOP- and Be-coated converters at $t=800~\mathrm{s}$, near saturation. The maximum initial UCN energy was set to $233.5~\mathrm{neV}$ for both converter wall materials.}
\label{final}
\end{figure}

Figure~\ref{final} shows the UCN energy spectra within the Be- and CYTOP-coated converters at $t=800~\mathrm{s}$, when the UCN populations approach saturation. Despite identical UCN input conditions, the lower optical potential of CYTOP results in a smaller population of stored UCNs than that of Be. The simulation results indicate that the total storage time constants are 128~s and 133~s for the Be- and CYTOP-coated converters, respectively. The corresponding saturated UCN densities are 315 and 90~UCN/cm$^3$, respectively. With extraction efficiencies of 84\% for Be and 66\% for CYTOP, approximately $1.3\times10^6$ and $2.8\times10^5$ UCNs can be extracted per cycle, respectively. The predicted performance metrics are summarized in Table~\ref{tab:results}.

\begin{table*}[htbp]
\centering
\caption{Predicted performance of the preliminary superfluid $^{4}\mathrm{He}$ UCN source design at CSNS.}
\label{tab:results}
\begin{tabular}{lcc}
\toprule
\textbf{Converter wall material} & \textbf{Be} & \textbf{CYTOP} \\
\midrule
Total storage time constant (\text{s}) & 128 & 133 \\
Saturated UCN density (\text{UCN/cm$^3$}) & 315 & 90 \\
UCN production rate (\text{UCN/s}) & 12055 & 3185 \\
Extracted UCNs per cycle (UCNs, accumulation mode) & $1.3 \times 10^6$ & $2.8 \times 10^5$ \\
Extraction efficiency (\%) & 84 & 66 \\
\bottomrule
\end{tabular}
\end{table*}

\section{Conclusion}

In this study, a comprehensive analysis of the physical parameters governing the performance of the UCN source at CSNS has been conducted. We systematically evaluated the factors related to the incident cold neutrons, those governing the converter system, and those determining the extraction efficiency. Based on these optimizations, a preliminary physical design of the superfluid $^{4}\mathrm{He}$ UCN source at CSNS has been established. With a \SI{100}{\cm} (length) $\times$ \SI{7}{\cm} (width) $\times$ \SI{7}{\cm} (height) converter operating at \SI{0.7}{\kelvin}, the system is projected to achieve a saturated UCN density of \SI{315}{UCN/cm^3}, a total UCN production rate of \SI{12055}{UCN/s}, and an extraction efficiency of 84\% using a Be-coated converter, yielding approximately $1.3 \times 10^6$ extracted UCNs per cycle in accumulation mode. 

The physical design of the superfluid $^{4}\mathrm{He}$ UCN source presented in this study forms part of the overall physical design of the Neutron Physics and Application Spectrometer at CSNS, which successfully passed its physical design review on March 25, 2025. The spectrometer project has since entered the engineering design and construction stage.

\section{Acknowledgements}

This work was supported by the National Key R\&D Program of China (No. 2024YFE0110001), the National Natural Science Foundation of China (U1932219), and the Mobility 
Programme endorsed by the Joint Committee of the Sino-German Center 
(M-0728). The authors thank Dr. Geza Zsigmond for assistance in understanding the original MCUCN code and Prof. Oliver Zimmer for guidance on the SUN-series UCN sources. Special thanks to Prof. Kent K. Leung for providing the loss factor of dPS and to Dr. Andreas Frei for providing the loss factor of Replica. We thank Dr. Tiancheng Yi for his suggestions on improving the figures and for his comments on the first draft of the manuscript.

\end{document}